\documentclass[aps, prd, floatfix, nofootinbib, superscriptaddress, twocolumn]{revtex4-1}

\usepackage{latexsym}
\usepackage{amsmath}
\usepackage{amssymb}
\usepackage{amsfonts}
\usepackage{textcomp}
\usepackage{color}
\usepackage{CJKutf8}

\usepackage[mathscr,scaled=1.15]{urwchancal}
\DeclareFontFamily{OT1}{pzc}{}
\DeclareFontShape{OT1}{pzc}{m}{it}%
{<-> s * [1.15] pzcmi7t}{}
\DeclareMathAlphabet{\mathpzc}{OT1}{pzc}{m}{it}

\usepackage{color}

\usepackage{supertabular}
\usepackage{placeins}
\usepackage{epsfig}
\usepackage{graphicx}

\definecolor{purple}{rgb}{0.5,0,0.5}
\definecolor{blue}{rgb}{0.0,0,0.9}
\definecolor{prdblue}{rgb}{0.133,0.118,0.498}
\usepackage[colorlinks=true, pdfstartview=FitV, linkcolor=prdblue, citecolor= prdblue, urlcolor=prdblue]{hyperref}

\makeatletter

\setbox0\hbox{$\xdef\scriptratio{\strip@pt\dimexpr
    \numexpr(\sf@size*65536)/\f@size sp}$}

\newcommand{\scriptveryshortarrow}[1][3pt]{{%
    \hbox{\rule[\scriptratio\dimexpr\fontdimen22\textfont2-.2pt\relax]
               {\scriptratio\dimexpr#1\relax}{\scriptratio\dimexpr.4pt\relax}}%
   \mkern-4mu\hbox{\let\f@size\sf@size\usefont{U}{lasy}{m}{n}\symbol{41}}}}

\makeatother

\begin{document}

\begin{CJK}{UTF8}{song}

%Title of paper
%\title{Symmetry-preserving calculation of the pion's valence-quark distribution}
%\title{Unification of continuum and lattice predictions of the pion's valence distribution}
%\title{Symmetry, symmetry breaking, and pion parton distributions}
\title{$\,$\\[-6ex]\hspace*{\fill}{\normalsize{\sf\emph{Preprint no}.\ NJU-INP 054/22}}\\[1ex]
%E615}
Emergence of pion parton distributions}
%Informing analyses of pion parton distributions}

\date{2022 April 22}
%\date{2022 March 07}
%\date{2022 January 04}
%\date{2021 December 17}
%
\author{Z.-F.~Cui} %(崔著钫)}
%\email[]{phycui@nju.edu.cn}
\affiliation{School of Physics, Nanjing University, Nanjing, Jiangsu 210093, China}
\affiliation{Institute for Nonperturbative Physics, Nanjing University, Nanjing, Jiangsu 210093, China}
%--
\author{M.~Ding} %(丁明慧)}
%\email[]{m.ding@hzdr.de}
\affiliation{Helmholtz-Zentrum Dresden-Rossendorf, Bautzner Landstra{\ss}e 400, D-01328 Dresden, Germany}
\author{J. M. Morgado}
%\email[]{josemanuel.morgado@dci.uhu.es}
\affiliation{Department of Integrated Sciences and Center for Advanced Studies in Physics, Mathematics and Computation, University of Huelva, E-21071 Huelva, Spain.}
\author{K.~Raya}
%\email[]{khepani@ugr.es}
\affiliation{Departamento de F\'isica Te\'orica y del Cosmos, Universidad de Granada, E-18071, Granada, Spain}
\affiliation{Instituto de Ciencias Nucleares, Universidad Nacional Aut\'onoma de M\'exico, Apartado Postal 70-543, CDMX 04510, M\'exico}
%--
\author{D.~Binosi}
\email{binosi@ectstar.eu}
\affiliation{European Centre for Theoretical Studies in Nuclear Physics
and Related Areas, Villa Tambosi, Strada delle Tabarelle 286, I-38123 Villazzano (TN), Italy}
\author{\\L.~Chang} % (常雷)}
\email[]{leichang@nankai.edu.cn}
\affiliation{School of Physics, Nankai University, Tianjin 300071, China}
\author{F.~De Soto}
%--
%\email[]{fcsotbor@upo.es}
%\email[]{liulangtian@nju.edu.cn, phycui@nju.edu.cn, m.ding@hzdr.de, khepani@ugr.es,\\ binosi@ectstar.eu, leichang@nankai.edu.cn,\\ cdroberts@nju.edu.cn, jose.rodriguez@dfaie.uhu.es,\\ s.schmidt@hzdr.de}
\affiliation{Dpto.\ Sistemas F\'isicos, Qu\'imicos y Naturales, Univ.\ Pablo de Olavide, E-41013 Sevilla, Spain}
%--
\author{C.\,D.~Roberts}
\email[]{cdroberts@nju.edu.cn}
%\email[]{liulangtian@nju.edu.cn, phycui@nju.edu.cn, m.ding@hzdr.de, khepani@ugr.es,\\ binosi@ectstar.eu, leichang@nankai.edu.cn,\\ cdroberts@nju.edu.cn, jose.rodriguez@dfaie.uhu.es,\\ s.schmidt@hzdr.de}
\affiliation{School of Physics, Nanjing University, Nanjing, Jiangsu 210093, China}
\affiliation{Institute for Nonperturbative Physics, Nanjing University, Nanjing, Jiangsu 210093, China}
\author{J.~Rodr\'{\i}guez-Quintero}
\email[]{jose.rodriguez@dfaie.uhu.es}
\affiliation{Department of Integrated Sciences and Center for Advanced Studies in Physics, Mathematics and Computation, University of Huelva, E-21071 Huelva, Spain.}
%--
\author{S.\,M.~Schmidt}
\email[]{s.schmidt@hzdr.de}
\affiliation{Helmholtz-Zentrum Dresden-Rossendorf, Bautzner Landstra{\ss}e 400, D-01328 Dresden, Germany}
\affiliation{RWTH Aachen University, III. Physikalisches Institut B, Aachen D-52074, Germany}

\begin{abstract}
Supposing only that there is an effective charge which defines an evolution scheme for parton distribution functions (DFs) that is all-orders exact, strict lower and upper bounds on all Mellin moments of the valence-quark DFs of pion-like systems are derived.  Exploiting contemporary results from numerical simulations of lattice-regularised quantum chromodynamics (QCD) that are consistent with these bounds, parameter-free predictions for pion valence, glue, and sea DFs are obtained.  The form of the valence-quark DF at large values of the light-front momentum fraction is consistent with predictions derived using the QCD-prescribed behaviour of the pion wave function.
%\\
%
%\rule{0.78\linewidth}{0.1ex}
\end{abstract}
%%
%%Keywords:
%%proton \sep
%%magnetic charge radius \sep
%%electric charge radius \sep
%%emergence of mass \sep
%%lepton-hadron scattering \sep
%%strong interactions in the standard model of particle physics

\maketitle

\end{CJK}

%%%%%%%%%%%%%%%%%%%%%%%%%%%%%%%%%%%%%%%%%%%%%%%%%%%%%%%%%%%%%%%%%%%%%%%%%%%%%%%%%%%%%%%%%%%%%%%%%%%%%%%%%%%%%%%%%%%%%%%
% 4500 words

%\noindent\emph{1.$\;$Issues and motivations} ---
\section{Issues and motivations}
Within the Standard Model of particle physics, hadrons emerged roughly $1\,\mu$s after the Big Bang \cite{Boyanovsky:2006bf}.  At this time, the colour-carrying gluon and quark (parton) degrees-of-freedom, in terms of which the Lagrangian of quantum chromodynamics (QCD) is expressed, were sublimated into colour-singlet bound states with nuclear-size masses and femtometre-scale radii.  Pions ($\pi^\pm$, $\pi^0$) are the lightest hadrons; and without them, even simple nuclei could not have formed in the ensuing few minutes \cite[Sec.\,24]{Zyla:2020zbs}.  Additionally, and crucially for the stability of nuclei, pions are unnaturally light: compared with the masses of the protons and neutrons ($m_N$) they bind, $m_\pi \approx 0.15 m_N$.  This is explained if the pions are Nambu-Goldstone (NG) bosons associated with dynamical chiral symmetry breaking in QCD \cite{Lane:1974he, Politzer:1976tv, Marciano:1977su}.  That raises some very basic questions, \emph{e.g}.: what imprints, if any, does this NG boson character leave on pion structure; and does it distinguish their structure and interactions from those of the nucleons they bind?

\begin{figure}[b]
\centerline{\includegraphics[width=0.27\textwidth]{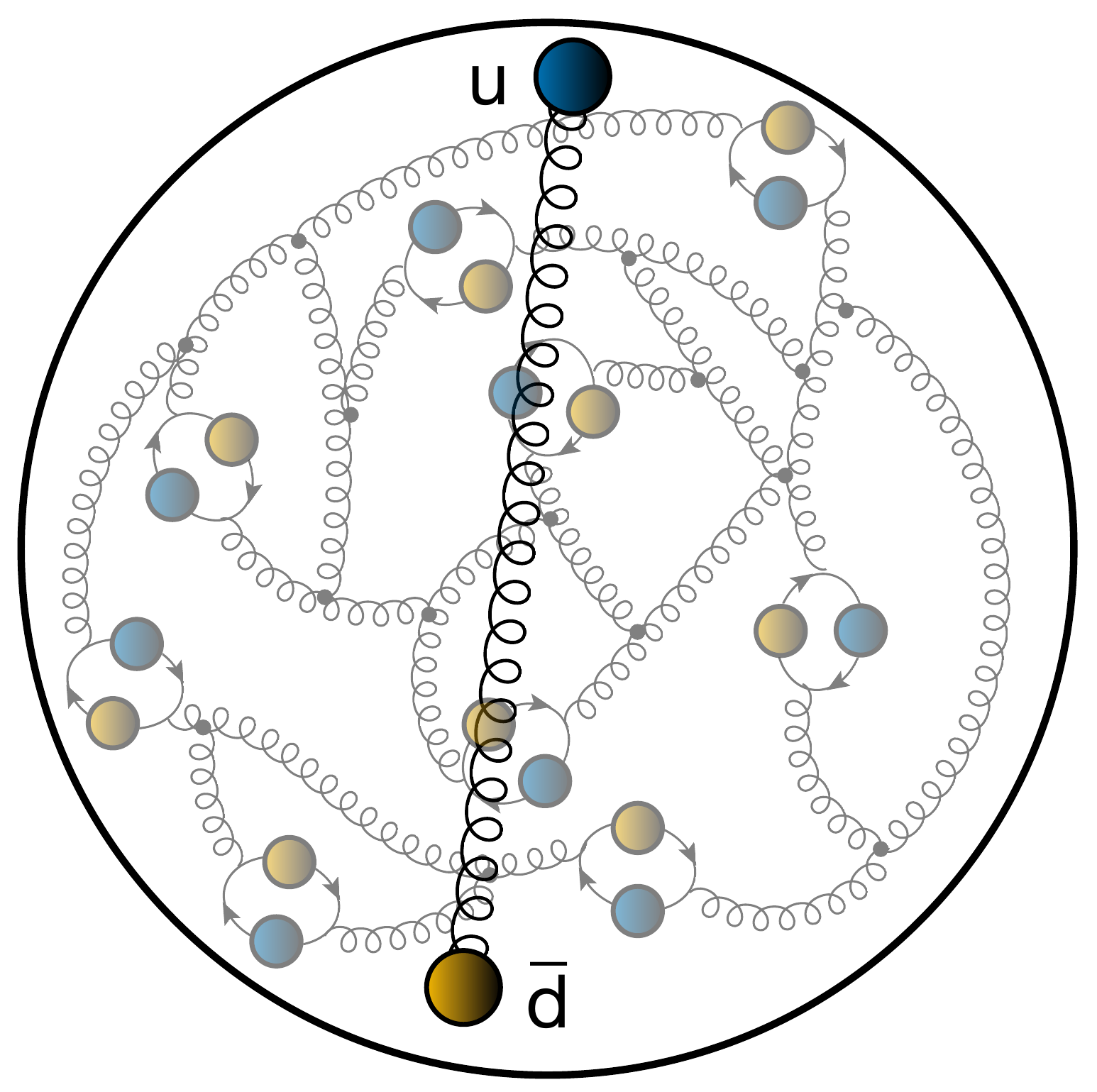}}
\caption{\label{ImagePion}
In terms of QCD's Lagrangian degrees-of-freedom, the $\pi^+$ contains one valence $u$-quark, one valence $\bar d$-quark, and, owing to the strong-interaction, infinitely many gluons and sea quarks, indicated here as ``springs'' and closed loops, respectively.
($\pi^-$ is $d\bar u$ and $\pi^0$ is $u \bar u - d\bar d$.)
}
\end{figure}

In QCD, pions are bound-states seeded by a valence-quark and valence-antiquark, Fig.\,\ref{ImagePion}.
%: $\pi^+ \sim u\bar d$, $\pi^-\sim d\bar u$, $\pi^0 \sim u\bar u - d\bar d$.
Yet, their properties cannot be determined by solving a typical two-body problem in quantum mechanics.  Owing to strong self-interactions amongst gluons -- QCD's gauge bosons, the Lagrangian gluon and quark partons are transmogrified into complex quasiparticles.  Each parton species evolves to acquire a distinct dynamically generated running mass \cite{Aguilar:2008xm, Gao:2017uox, Roberts:2021nhw}, both of which are large at infrared momenta and typified by a renormalisation group invariant mass $m_0\approx m_N/2$; and the interactions between these quasiparticles are described by a momentum dependent coupling \cite{Binosi:2016nme, Cui:2019dwv}, $\hat\alpha(k^2)$, which runs to saturate at infrared momenta: $\hat\alpha(k^2\lesssim m_0^2)\approx \pi$.  These features are primary signals of the dynamical breaking of scale invariance in QCD \cite{Roberts:2016vyn}, \emph{i.e}., the phenomenon of emergent hadron mass (EHM).  They produce a pion whose structure, when unfolded in terms of parton degrees-of-freedom, has the complicated character illustrated in Fig.\,\ref{ImagePion}, and may lead to gluon and quark confinement \cite{Krein:1990sf, Binosi:2019ecz}.

Continuum Schwinger function methods (CSMs) \cite{Eichmann:2016yit, Fischer:2018sdj, Qin:2020rad, Roberts:2020hiw} are well suited to tackling the pion.  Successes have been achieved by solving the coupled quark gap and meson Bethe-Salpeter equations to obtain the pion's Bethe-Salpeter wave function, $\chi_\pi(k,k-P)$, where $P$ is the pion's total momentum and $k$ is the momentum of the valence-quark, and exploiting that to predict pion observables \cite{Horn:2016rip, Roberts:2021nhw}.  Of great importance to an explanation of pion properties, is the expression of an intimate link between the dressed-quark mass function and $\chi_\pi$ \cite{Maris:1997tm, Holl:2004fr, Qin:2014vya, Williams:2015cvx, Binosi:2016rxz, Qin:2020jig}.
This link means that, although it can be studied using a large array of reactions \cite{Carman:2020qmb, Brodsky:2020vco, Barabanov:2020jvn, Roberts:2020hiw},
the sharpest probes of EHM are  found in pion properties; raising pion structure studies to the highest level of importance \cite{JlabTDIS1, JlabTDIS2, Adams:2018pwt, Aguilar:2019teb, Chen:2020ijn, Anderle:2021wcy, Arrington:2021biu}.

Much is promised by data relating to pion parton distribution functions (DFs), \emph{viz}.\ the probability densities describing the light-front momentum fractions carried by each parton species within the pion \cite{Holt:2010vj}.
%and, more generally, within any hadron \cite{Holt:2010vj}.
For instance, ${\mathpzc u}^\pi(x;\zeta)$ is the density for finding a valence $u$-quark with momentum fraction $x$ when the pion is resolved at scale $\zeta$.  On $\zeta \lesssim 2 m_N$, this valence-quark is not equivalent to a valence-quark-parton; rather, it is connected to that parton as an object dressed by interactions in the manner described by the quark gap equation \cite{Roberts:1994dr}.  Undressing reveals the complexities in Fig.\,\ref{ImagePion}, leading to growth of the glue and sea-quark DFs, ${\mathpzc g}^\pi(x)$, ${\mathpzc S}^\pi(x)$.  However, more than forty years after the first experiment to collect data suitable for extracting pion DFs \cite{Corden:1980xf, Badier:1983mj, Betev:1985pg, Conway:1989fs}, the behaviour of all these functions remains uncertain and controversial \cite{Chang:2021utv, Cui:2021mom}: some analyses potentially challenge QCD as the theory of strong interactions.  New experiments \cite{JlabTDIS1, JlabTDIS2, Adams:2018pwt, Aguilar:2019teb, Chen:2020ijn, Anderle:2021wcy, Arrington:2021biu} will hopefully serve to dispel the confusion.

%\smallskip

%\noindent\emph{2.$\;$Symmetry and pion wave functions} ---
\section{Symmetry and pion wave functions}
Notwithstanding the intricacies of Fig.\,\ref{ImagePion}, %a structural
simplicity emerges when one adapts CSMs to the pion problem.  Then, at infrared scales, the $\pi^+$, for instance, appears as a two-body bound-state of a dressed-valence-quark, ${u}$, and a dressed-valence-antiquark $\bar {d}$, with the complexity hidden from view because the infinitely many gluon and quark partons have been absorbed into making the dressed quasiparticles. %degrees-of-freedom.
%Well-defined mathematical operations generate this outcome \cite{Horn:2016rip, Roberts:2021nhw}.
In this case, exploiting the ${\cal G}$-parity symmetry limit, which is an accurate reflection of Nature,
\begin{equation}
\label{symmetry}
\chi_\pi(k,k-P) = \chi_\pi(-k+P,-k)\,.
\end{equation}
%%% changing sign of relative momentum.
%$\bar {\mathpzc d}^\pi(x;\zeta)={\mathpzc u}^\pi(x;\zeta)$.

Unlike wave functions in quantum mechanics, $\chi_\pi$ does not have a probability interpretation; hence, cannot directly yield ${\mathpzc u}^\pi(x;\zeta)$.  That door is opened by projection to obtain the associated light-front wave function (LFWF) \cite{tHooft:1974pnl, Chang:2013pq}, $\psi_\pi(x,|\vec{k}_\perp|^2;\zeta)$, which is a probability amplitude.
Here, using linearly independent four-vectors $n$, $\bar n$, with $n^2=0=\bar n^2$, $n\cdot \bar n=-1$:
$x=n\cdot k/n\cdot P$, \emph{i.e}., the light-front fraction of the pion's total momentum carried by the valence-quark;
and
$\vec{k}_\perp$
%, the two-vector built from the in-general nonzero components of
%$k_\mu = O^\perp_{\mu\nu}k_\nu$, $ O^\perp_{\mu\nu} = \delta_{\mu\nu} + n_\mu \bar n_\nu + \bar n_\mu n_\nu$, \emph{viz}.\
is that part of the valence-quark's momentum which lies in the light-front transverse plane.

Using the LFWF,
\begin{equation}
\label{Eqvalence}
{\mathpzc u}^\pi(x;\zeta) \stackrel{x\in(0,1)}= H_\pi^{\mathpzc u}(x,t=0;\zeta)\,,
\end{equation}
where $H_\pi^{\mathpzc u}$ is the valence $u$-quark forward generalised parton distribution \cite{Diehl:2000xz}, and:
\begin{align}
H_\pi^{\mathpzc u}(x,0;\zeta)
&= \int \frac{d^2{k_\perp}}{16 \pi^3}
|\psi_{\pi}^{u}\left( x,{k}_{\perp}^2;\zeta \right)|^2 \,.
%\psi_{\pi}^{u\ast}\left(x,{k}_{\perp}^2;\zeta \right)
%\nonumber \\ && \rule[0cm]{3.25cm}{0cm} \times
%\psi_{\pi}^{u}\left( x,{k}_{\perp}^2;\zeta \right) \,.
\label{eq:overlap}
\end{align}
%where $k_{\perp \pm}=k_\perp \pm (1-x)t/2$, with $t$ the squared mo\-men\-tum-transfer to the $\pi$ in processes designed to measure the GPD.
%
The LFWF defined by projection of $\chi_\pi(k,k-P)$ is associated with a scale, $\zeta=\zeta_{\cal H}$, at which the dressed-quark and -antiquark carry all pion properties and Eq.\,\eqref{symmetry} entails
%\begin{equation}
$\psi_\pi^u(x,|\vec{k}_\perp|^2;\zeta_{\cal H}) = \psi_\pi^u(1-x,|\vec{k}_\perp|^2;\zeta_{\cal H})$.
% \,.
%\end{equation}
Hence,
\begin{align}
\label{Equpisymmetry}
{\mathpzc u}^\pi(x;\zeta_{\cal H}) & = {\mathpzc u}^\pi(1-x;\zeta_{\cal H}) \,,\\
\label{twoxone}
\langle 2 x \rangle_{{\mathpzc u}_\pi}^{\zeta_{\cal H}} & := \int_0^1 dx\, 2 x {\mathpzc u}^\pi(x;\zeta_{\cal H})\rangle = 1\,,
\end{align}
confirming that dressed valence degrees-of-freedom carry all the pion's light-front momentum at this scale.  Momentum conservation demands that the glue and sea momentum fractions vanish at $\zeta_{\cal H}$; and since DFs are nonnegative on $x\in[0,1]$, then ${\mathpzc g}^\pi(x;\zeta_{\cal H})\equiv 0 \equiv {\mathpzc S}^\pi(x;\zeta_{\cal H})$.

As the resolving scale is increased to $\zeta>\zeta_{\cal H}$, the dressed-quark and -antiquark begin to shed their clothing, gluon emission and subsequent splitting commence \cite{Brodsky:1979gy}, and QCD evolution (DGLAP) \cite{Dokshitzer:1977sg, Gribov:1971zn, Lipatov:1974qm, Altarelli:1977zs} proceeds to generate nonzero glue and sea distributions from the nonperturbative information contained in ${\mathpzc u}^\pi(x;\zeta_{\cal H})$.  Thus, the complex structure in Fig.\,\ref{ImagePion} emerges.
%Consequently, when analysing data, one should not independently fit valence-quark, glue, and sea distributions at any scale $\zeta>\zeta_{\cal H}$ because the glue and sea distributions at $\zeta > \zeta_{\cal H}$ are completely determined by ${\mathpzc u}^\pi(x;\zeta_{\cal H})$ and the evolution equations: the glue and sea distributions are not independent functions.

A prediction for the value of $\zeta_{\cal H}$ follows from the properties of QCD's renormalisation group invariant effective charge \cite{Cui:2019dwv, Ding:2019qlr, Ding:2019lwe, Cui:2020dlm, Cui:2020tdf, Chang:2021utv}, $\hat\alpha(k^2)$.  Its scale is set by $m_0$, the gluon mass \cite{Cui:2019dwv, Roberts:2021nhw}.   Notwithstanding that, the value of $\zeta_{\cal H}$ is immaterial herein, so long as Eq.\,\eqref{Equpisymmetry} is understood.

Introducing the distribution ${\mathpzc P}(t) = {\mathpzc u}^\pi([1+t]/2;\zeta_{\cal H})$, the Mellin moments of the pion valence-quark DF are:
%\begin{subequations}
\begin{align}
%%\langle x^n & \rangle_{{\mathpzc u}_\pi}^{\zeta_{\cal H}}
%%= \frac{1}{2^{n+1}}\int_{-1}^{1} dt\,(1+t)^n {\mathpzc P}(t)\\
%
\langle x^n\rangle_{{\mathpzc u}_\pi}^{\zeta_{\cal H}}  &  = \frac{1}{2^n}\sum_{i=0}^{[n/2]}
\left(\begin{array}{c}
n\\2i\end{array}\right)\langle t^{2i}\rangle_{\mathpzc P}\,, %\\
%
%\langle t^{j}\rangle_{\mathpzc P} & = \int_0^1dt\,t^{j}{\mathpzc P}(t)\,.
%\int_0^1dt\,t^{2i}{\mathpzc P}(t)\,,\\
%
%\langle t^{2i}\rangle_{\mathpzc P} & = \int_0^1dt\,t^{2i}{\mathpzc P}(t)\,,
\end{align}
%\end{subequations}
%
$\langle t^{j}\rangle_{\mathpzc P}  = \int_0^1dt\,t^{j}{\mathpzc P}(t)$.
Since the hadron scale DF of a ground-state pseudoscalar meson is necessarily unimodal \cite[Sec.\,3]{Roberts:2021nhw}, two limiting cases are apparent:
(\emph{i}) ${\mathpzc P}(t) = \delta(t)$, corresponding to a pion constituted from two infinitely-massive valence constituents;
and
(\emph{ii}) its antithesis, ${\mathpzc P}(t) = \theta(1+t)\theta(1-t)$,
%where $\theta(\tau)$ is the Heaviside function,
which is obtained for a massless pion using a symmetry-preserving treatment of a vector$\,\times\,$vector contact interaction \cite{Zhang:2020ecj}.
They lead to the following bounds:
\begin{equation}
\label{momentbounds}
\frac{1}{2^n} \stackrel{\rm(\emph{i})}{\leq}
\langle x^n\rangle_{{\mathpzc u}_\pi}^{\zeta_{\cal H}}
\stackrel{\rm(\emph{ii})}{\leq} \frac{1}{1+n}\,.
\end{equation}

%\smallskip

%\noindent\emph{3.$\;$Principle and practice of all-orders evolution} ---
%
\section{Principle and practice of all-orders evolution}
We proceed by exploring the consequences of the following hypothesis \cite{Cui:2021mom}:\\[0.5ex]
\hspace*{1.5em}\parbox[t]{0.9\linewidth}{{\sf P1} -- \emph{There exists at least one effective charge, $\alpha_{1\ell}(k^2)$, such that, when used to integrate the one-loop DGLAP equations, an evolution scheme for parton DFs is defined that is all-orders exact}.}  %$\alpha_{1\ell}(k^2)$ need not be unique.
\smallskip

\noindent Charges of this type are discussed in Refs.\,\cite{Grunberg:1982fw, Grunberg:1989xf, Dokshitzer:1998nz}.  They need not be process-independent (PI); hence, not unique.
%: different charges may be needed for distinct observables.
Nevertheless, a suitable PI charge is not excluded, \emph{e.g}., that discussed in %Refs.\,\cite{Cui:2019dwv, Cui:2020dlm, Cui:2020tdf, Chang:2021utv}
Refs.\,\cite{Cui:2019dwv, Cui:2020tdf}
has proved efficacious.  In being defined via an observable -- in this case, pion structure functions, each such $\alpha_{1\ell}(k^2)$ is \cite{Deur:2016tte}: consistent with the renormalisation group; renormalisation scheme independent; everywhere analytic and finite; and supplies an infrared completion of any standard running coupling.

Regarding this hypothesis, it is worth observing here that CSM results for pion $\zeta=\zeta_{\cal H}$ valence DFs, obtained from symmetry-preserving analyses and used as initial values for evolution according to {\sf P1}, yield predictions for all pion $\zeta > \zeta_{\cal H}$ DFs (valence, sea, glue) that are consistent with QCD expectations, including those on their small- and large-$x$ behaviour \cite{Brodsky:1994kg, Yuan:2003fs, Cui:2021mom}.  Owing to a deficit of pion data \cite[Table~9.5]{Roberts:2021nhw}, more cannot yet be said.
On the other hand, given the large amount of relevant proton data, one might think it possible to test a variant of {\sf P1} using phenomenological proton DF fits  \cite{NNPDF:2017mvq, Hou:2019efy}.  Unfortunately, however, extant such fits are inconsistent with a range of QCD constraints; so, they cannot serve as a reliable foundation for testing the validity of evolution schemes related to {\sf P1}.  In large part, this explains conclusions drawn elsewhere \cite{Diehl:2019fsz}.  Future such studies should be built upon improved DF fits and use an effective charge that furnishes an infrared completion of QCD.

%Numerous consequences of {\sf P1} are described elsewhere \cite{Raya:2021zrz, Cui:2021N}.  Herein, we reveal and exploit a novel corollary, which enables us to deliver parameter-free predictions for the pointwise behaviour of all pion DFs based solely on input from numerical simulations of lattice-regularised QCD \cite{Oehm:2018jvm, Sufian:2019bol, Joo:2019bzr, Alexandrou:2021mmi}.

\begin{table}[t!]
\caption{%\small
\label{latticemoments}
Lattice-QCD results for Mellin moments of the pion valence-quark DF at
$\zeta=\zeta_2=2\,$GeV \cite{Joo:2019bzr}
and
$\zeta_5=5.2\,$GeV \cite{Sufian:2019bol, Alexandrou:2021mmi}
%Oehm:2018jvm, Sufian:2019bol, Joo:2019bzr, Alexandrou:2021mmi
%% \mbox{lQCD\,\cite{Oehm:2018jvm}} & 0.21(1) & 0.16(3) & \\
%% \mbox{lQCD\,\cite{Joo:2019bzr}} & 0.254(03) & 0.094(12) & 0.057(04) \\\hline
%% 0.18(3) & 0.064(10) & 0.030(5)\\
}
%% n l1 l2 l3 l4
\begin{tabular*}%{|c|c|c|c|c|c|c|}\hline
{\hsize}
{
l@{\extracolsep{0ptplus1fil}}
|l@{\extracolsep{0ptplus1fil}}
l@{\extracolsep{0ptplus1fil}}
l@{\extracolsep{0ptplus1fil}}
l@{\extracolsep{0ptplus1fil}}}\hline
%%\begin{tabular}{l|c|c|c|c|c|c|c}\hline
$n\ $ & \cite{Joo:2019bzr} & \cite{Sufian:2019bol} & \cite{Alexandrou:2021mmi} \\\hline
$1\ $ & $0.254(03)\ $ & $0.18(3)\ $& $0.23(3)(7)\ $\\
$2\ $ & $0.094(12)\ $ & $0.064(10)\ $& $0.087(05)(08)\ $\\
$3\ $ & $0.057(04)\ $ & $0.030(05)\ $& $0.041(05)(09)\ $\\
$4\ $ & &  & $0.023(05)(06)\ $\\
$5\ $ & &  & $0.014(04)(05)\ $\\
$6\ $ & &  & $0.009(03)(03)\ $\\\hline
\end{tabular*}
\end{table}

{\sf P1} entails \cite[Sec.\,VII]{Raya:2021zrz}
\begin{equation}
\label{EqxnzzH}
\langle x^n\rangle_{{\mathpzc u}_\pi}^\zeta
= \langle x^n\rangle_{{\mathpzc u}_\pi}^{\zeta_{\cal H}}
\left( \langle 2 x \rangle_{{\mathpzc u}_\pi}^\zeta\right)^{\gamma_0^n/\gamma_0^1},
\end{equation}
where $\gamma_0^0=0$ and, for $n_f=4$ quark flavours, $\gamma_0^{1,2}=32/9, 50/9$.  The higher-$n$ results are listed elsewhere \cite[Eq.\,(56a)]{Raya:2021zrz}.  Thus, given the pion valence-quark DF at one scale, \emph{e.g}., $\zeta_{\cal H}$, then its pointwise behaviour at any other scale, $\zeta$, is fully determined by the value of its first moment at $\zeta$.  No other knowledge is required; especially, one need know nothing about the actual form of $\alpha_{1\ell}(k^2)$.  Similar statements are true for ${\mathpzc g}^\pi(x;\zeta)$, ${\mathpzc S}^\pi(x;\zeta)$.  As noted above, the hadron scale is uniquely defined by $\langle 2 x \rangle_{{\mathpzc u}_\pi}^{\zeta_{\cal H}} = 1$.  Inserting Eq.\,\eqref{EqxnzzH} into Eq.\,\eqref{momentbounds}, one finds:
\begin{equation}
\label{momentbounds2}
\frac{1}{2^n}
\leq
\langle x^n\rangle_{{\mathpzc u}_\pi}^{\zeta}
(\langle 2 x\rangle_{{\mathpzc u}_\pi}^{\zeta})^{-\gamma_0^n/\gamma_0^1}
%(\langle 2 x\rangle_{{\mathpzc u}_\pi}^{\zeta})^{-\gamma_0^n/\gamma_0^1}
\leq \frac{1}{1+n}\,.
\end{equation}

Together, Eqs.\,\eqref{Equpisymmetry}, \eqref{EqxnzzH} entail this recursion \cite{Cui:2021mom}:
\begin{align}
\langle & x^{2 n+1}\rangle_{{\mathpzc u}_\pi}^{\zeta}
= \frac{(\langle 2 x \rangle_{{\mathpzc u}_\pi}^\zeta)^{\gamma_0^{2n+1}/\gamma_0^1}}{2(n+1)} \nonumber \\
& \times \sum_{j=0,1,\ldots}^{2n}(-)^j
\left(\begin{array}{c} 2(n+1) \\ j \end{array}\right)
\langle x^j\rangle_{{\mathpzc u}_\pi}^{\zeta}
(\langle 2 x \rangle_{{\mathpzc u}_\pi}^\zeta)^{-\gamma_0^{j}/\gamma_0^1}\,.
\label{EqSymmB}
\end{align}
%%
%\begin{align}
%\langle & x^{2 n+1}\rangle_{{\mathpzc u}_\pi}^{\zeta_{\cal H}}\nonumber  \\
%
%& = \frac{1}{2(n+1)}\sum_{j=0,1,\ldots}^{2n}(-)^j
%\left(\begin{array}{c} 2(n+1) \\ j \end{array}\right) \langle x^j\rangle_{{\mathpzc u}_\pi}^{\zeta_{\cal H}}\,.
%\label{EqSymmB}
%\end{align}
Namely, for any symmetric function, Eq.\,\eqref{Equpisymmetry}, which evolves according to Eq.\,\eqref{EqxnzzH}, the odd-order Mellin moment $\langle x^{2n+1} \rangle_{{\mathpzc u}_\pi}^{\zeta}$ is completely determined by the set of even moments $\langle x^{2m} \rangle_{{\mathpzc u}_\pi}^{\zeta}$ with $m\leq n$.  Conversely, if a DF satisfies Eq.\,\eqref{EqSymmB}, then it is linked by evolution to a symmetric distribution at $\zeta_{\cal H}$.

%\smallskip

%\noindent\emph{4.$\;$Pion valence-quark DF from lattice-QCD moments} ---
\section{Pion valence-quark DF from lattice-QCD moments}
\begin{figure}[t]
\centerline{\includegraphics[width=0.43\textwidth]{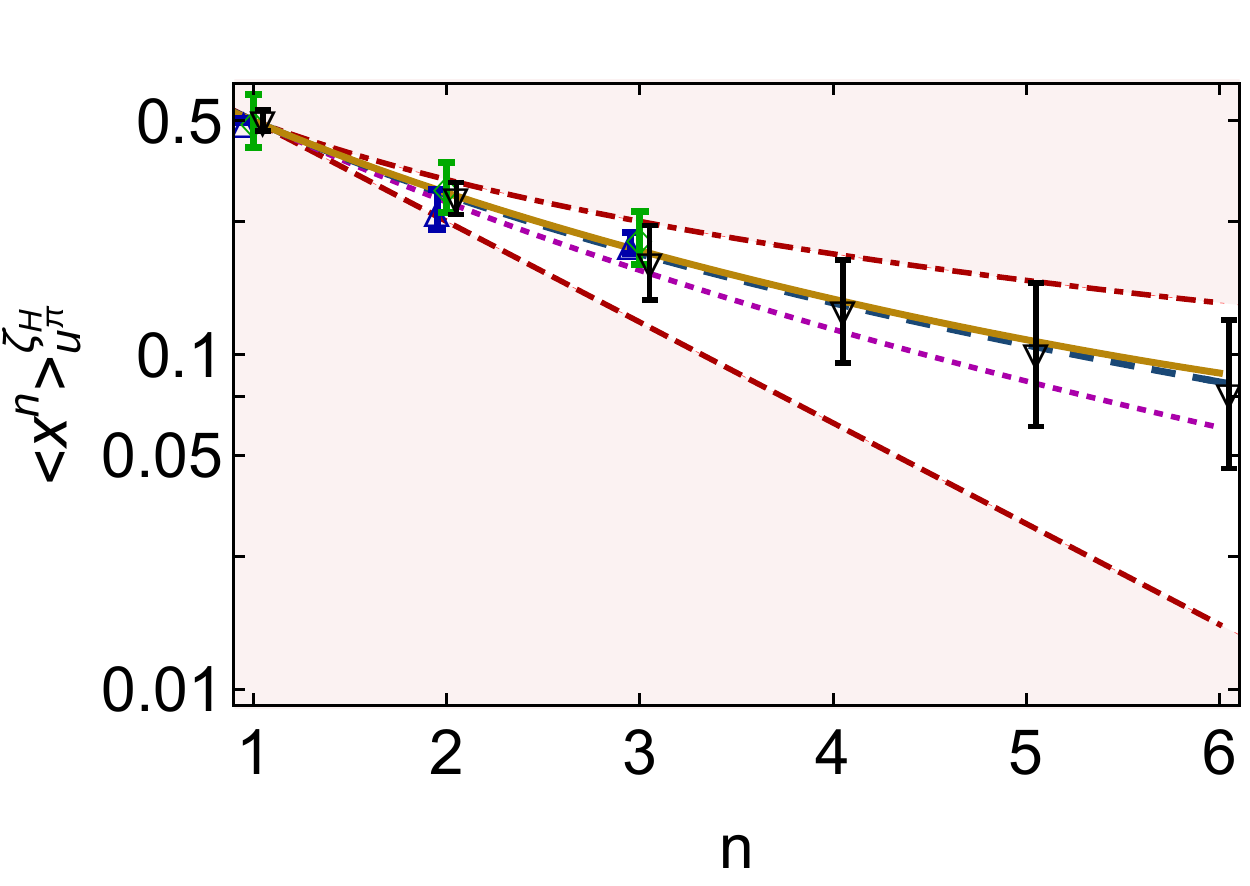}}
\caption{\label{PlotMoments}
Mellin moments from Table~\ref{latticemoments}, referred to $\zeta_{\cal H}$ via
Eq.\,\eqref{EqxnzzH}.
%%$\langle x^n\rangle_{{\mathpzc u}_\pi}^{\zeta}
%%(\langle 2 x\rangle_{{\mathpzc u}_\pi}^{\zeta})^{-\gamma_0^n/\gamma_0^1}$:
blue up-triangles \cite{Joo:2019bzr};
green diamonds \cite{Sufian:2019bol}; and
black down-triangles \cite{Alexandrou:2021mmi}.
Results consistent with the bounds in Eq.\,\eqref{momentbounds2} fall within the open band.  The excluded regions are lightly shaded in red.
Gold curve: trajectory of moments that minimises Eq.\,\eqref{X2distribution}.
Long-dashed dark-blue curve: moments of CSM distribution \cite{Cui:2020tdf}.
Dotted magenta curve: moments of the scale-free distribution: ${\mathpzc q}^{\rm sf}(x)=30 x^2(1-x)^2$.
}
\end{figure}

Recent years have seen the refinement of lattice-QCD predictions for low-order Mellin moments of the pion valence-quark DF.  Some contemporary results are listed in Table~\ref{latticemoments} and plotted in Fig.\,\ref{PlotMoments}.  They satisfy the bounds in Eq.\,\eqref{momentbounds2}.  Importantly, a calculation which yields points that lie systematically outside the inclusion area does not describe a physically realisable pion-like bound-state; or, stated otherwise, contains systematic uncertainties that preclude its connection with a physical pion-like system.

The moments in Table~\ref{latticemoments}--Column~3 \cite{Alexandrou:2021mmi} satisfy Eq.\,\eqref{EqSymmB}; hence, are associated with a symmetric pion valence-quark DF at $\zeta_{\cal H}$.  Here one sees that the moments in Refs.\,\cite{Joo:2019bzr, Sufian:2019bol} are compatible with those in Ref.\,\cite{Alexandrou:2021mmi}; so may also be associated with a symmetric DF at $\zeta_{\cal H}$.
Moreover, the consistency between the results in Table~\ref{latticemoments} means that one can combine the moments and seek an optimal description of the entire collection.

We therefore consider the symmetric distribution
\begin{equation}
\label{EqRefSymmetric}
%{\mathpzc u}^\pi(x;\zeta_{\cal H}) = {\mathpzc n}_0 x^\beta (1-x)^\beta [1-\sqrt{x(1-x)}]^2\,,
{\mathpzc u}^\pi(x;\zeta_{\cal H}) = {\mathpzc n}_0 \ln(1+x^2(1-x)^2/\rho^2)\,,
%Log[1 + x^2 (1 - x)^2 /\[Alpha]^2]/
\end{equation}
%%  alpha = 0.048 +/- 0.029,
${\mathpzc n}_0$ ensures unit normalisation, which is simple yet flexible enough to express the dilation that EHM is known to introduce \cite{Ding:2019qlr, Ding:2019lwe, Cui:2019dwv, Cui:2020dlm, Cui:2020tdf, Chang:2021utv}.
%
%$1/{\mathpzc n}_0=[(5 \beta +7) \Gamma (\beta +1) \Gamma (\beta +2)-2 (2 \beta +3) \Gamma \left(\beta +\tfrac{3}{2}\right)^2]/\Gamma (2 \beta +4)$,
%
Denoting the moments of this distribution by ${\mathpzc M}^\pi_n(\rho)$, we minimise the following uncertainty-weighted $\chi^2$-function:
\begin{equation}
\chi^2(\rho) = \!\!\!
\sum_{{\rm s}=\mbox{\tiny\cite{Joo:2019bzr, Sufian:2019bol, Alexandrou:2021mmi}}}
\sum_{n=2}^6 a^{\rm s}_n
\frac{ ({\mathpzc M}^\pi_n(\rho)-M^{\rm s}_n(\zeta)/(2 M^{\rm s}_1)^{\gamma_0^n/\gamma_0^1})^2}
{(\sigma^{\rm s}_n)^2}\,, \label{X2distribution}
\end{equation}
where $a^{\rm s}_n=1$ in all cases with an entry in Table~\ref{latticemoments} and is otherwise zero; and $M^{\rm s}_n(\zeta)$, $\sigma^{\rm s}_n$ are the related nonzero entries, \emph{viz}.\ moment and uncertainty.  This yields $\rho_0 = 0.048$ and $\chi^2(\rho_0)/\mbox{degree-of-freedom}=0.27$.
% 1.44278
% 0.0462374
The associated trajectory of moments is drawn in Fig.\,\ref{PlotMoments} (gold curve).
It is practically indistinguishable from that calculated using the CSM DF prediction \cite{Ding:2019qlr, Ding:2019lwe, Cui:2020dlm, Cui:2020tdf}.
(For subsequent use, we rescale the uncertainties in Eq.\,\eqref{X2distribution} such that $\chi_0^2:=\chi^2(\rho_0) = d-2$, where $d=8$ is the number of degrees-of-freedom.)

\begin{figure}[t]
\vspace*{2ex}

\leftline{\hspace*{0.5em}{\large{\textsf{A}}}}
\vspace*{-3ex}
\includegraphics[width=0.43\textwidth]{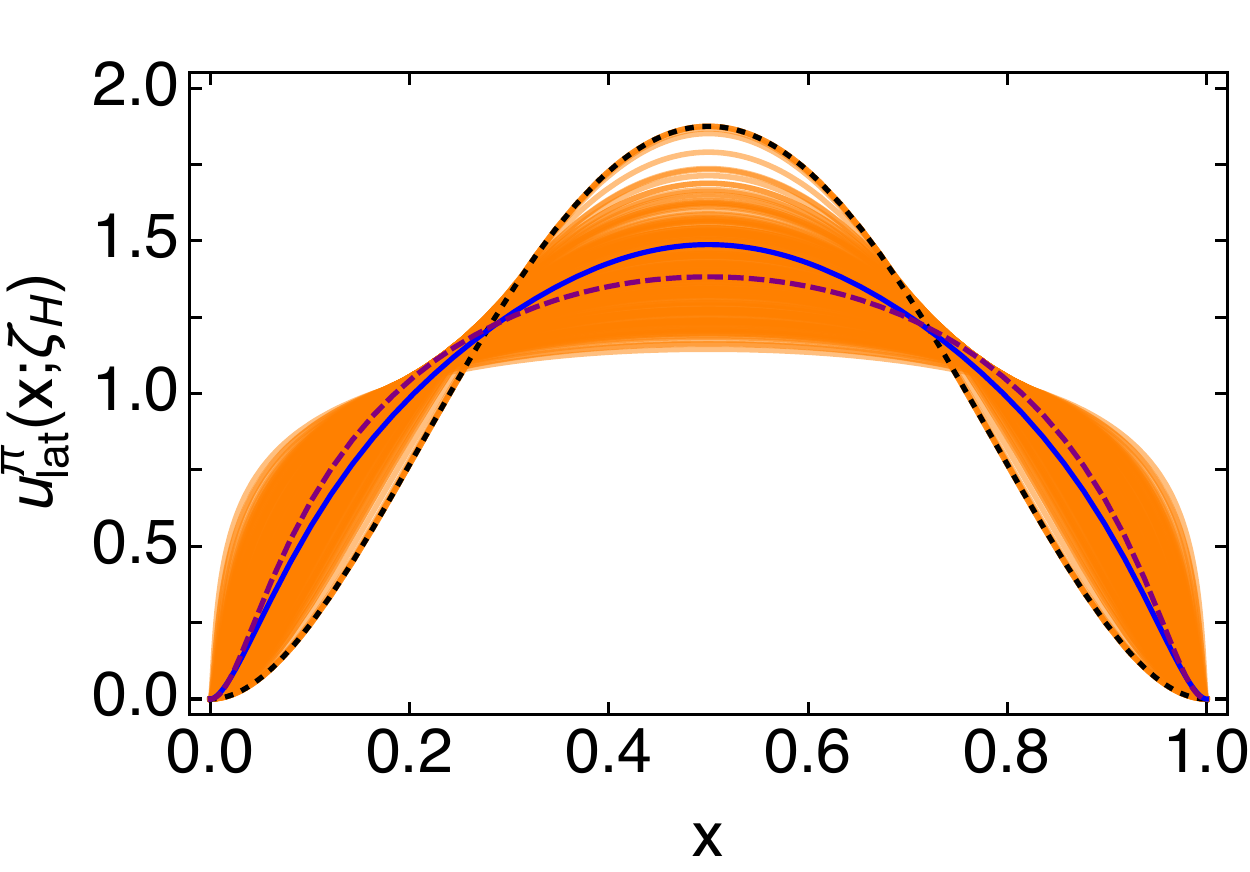}
\vspace*{-1ex}

\leftline{\hspace*{0.5em}{\large{\textsf{B}}}}
\vspace*{-3ex}
\includegraphics[width=0.43\textwidth]{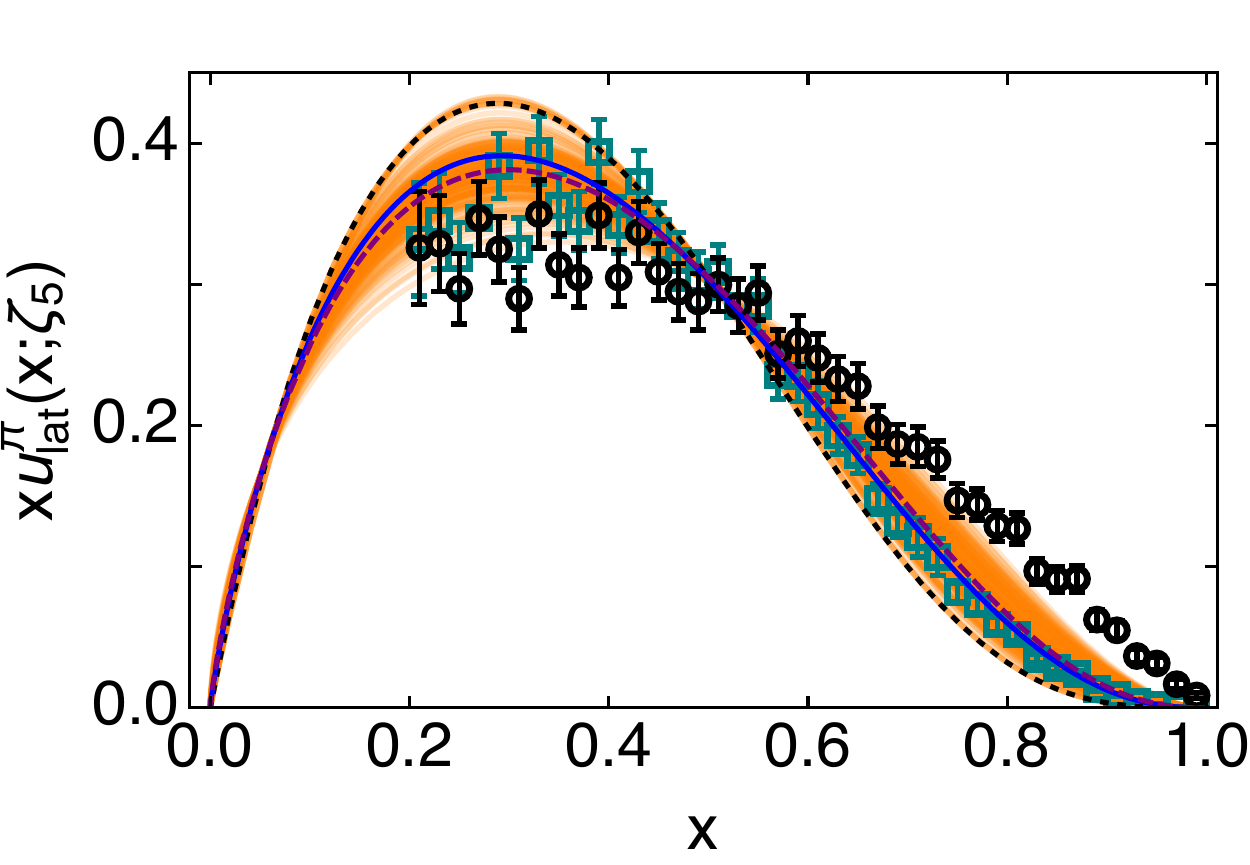}
\caption{\label{Fensemble}
\emph{Upper panel}\,--\,{\sf A}.  Randomly distributed ensemble of lattice-QCD-based \cite{Joo:2019bzr, Sufian:2019bol, Alexandrou:2021mmi} pion valence-quark DFs (orange curves) constructed using the procedure described in connection with Eq.\,\eqref{EqProb}.
\emph{Lower panel}\,--\,{\sf B}.  $\zeta_{\cal H}\to \zeta_{5}$ evolution of each curve in Panel {\sf A}. % (orange curves).
Black circles, data recorded in Ref.\,\cite[E615]{Conway:1989fs};
and teal boxes, reevaluation of that data as presented in Ref.\,\cite{Aicher:2010cb}.
Both panels.
Dashed magenta curve: central $\rho=\rho_0$ result in Eq.\,\eqref{EqRefSymmetric}.
Solid blue curve: CSM prediction from Refs.\,\cite{Cui:2020dlm, Cui:2020tdf, Chang:2021utv}.
Dotted black curve: scale-free distribution.  (All at scale appropriate to panel.)
}
\end{figure}

Based on this result, we generate an ensemble of curves that express the uncertainty in the lattice moments as follows.
(\emph{i}) From a distribution centred on $\rho_0$, choose a new value of $\rho$.
(\emph{ii})
Evaluate $\chi^2(\rho)$ in Eq.\,\eqref{X2distribution}.
The new value of $\rho$ is accepted with probability
\begin{equation}
\label{EqProb}
{\mathpzc P}  = \frac{P(\chi^2;d)}{P(\chi_0^2;d)} \,, \;
P(y;d) = \frac{(1/2)^{d/2}}{\Gamma(d/2)} y^{d/2-1} {\rm e}^{-y/2}\,.
\end{equation}
%%\begin{subequations}
%%\label{EqProb}
%%\begin{align}
%%{\mathpzc P} & = \frac{P(\chi^2;d)}{P(\chi_0^2;d)} \,, %  \leq 1\,,\\
%
%%P(y;d) & = \frac{(1/2)^{d/2}}{\Gamma(d/2)} y^{d/2-1} {\rm e}^{-y/2}\,,
%%\end{align}
%%\end{subequations}
%where $d=8$ and $\chi_0^2 = d-2$ locates the maximum of the $\chi^2$-probability density, $P(\chi^2;d)$.
%
(\emph{iii}) Repeat (\emph{i}) and (\emph{ii}) until one has a $K \gtrsim 200$-member ensemble of DFs.
This yields the DFs drawn in Fig.\,\ref{Fensemble}A.

Exploiting {\sf P1}, every curve in Fig.\,\ref{Fensemble}A can be evolved to $\zeta_5=5.2\,$GeV once $\langle 2 x\rangle_{{\mathpzc u}_\pi}^{\zeta_5}$ is known.  Using an uncertainty weighted average of the results in Refs.\,\cite{Cui:2020tdf, Joo:2019bzr, Sufian:2019bol, Alexandrou:2021mmi}, which yields $\langle 2 x\rangle_{{\mathpzc u}_\pi}^{\zeta_5}=0.435(12)$,
%% <2 x> @ \[Zeta]5 = 0.434919 +/- 0.0118778
and no additional information, one obtains the orange curves in Fig.\,\ref{Fensemble}B.
%Best-fit parameters at zeta5: N0=2.22(48), alpha=-0.17(7), beta=2.43(36), d=1.24(74)
The central curve and associated $1\sigma$-band are reproduced by
\begin{equation}
\label{lQCDreconstruct}
{\mathpzc u}^\pi(x;\zeta_5)
= {\mathpzc n}_0^{\zeta_5} x^\alpha (1-x)^\beta (1+\gamma x^2)\,,
\end{equation}
%$\alpha=-0.17_{\mp 0.07}$, $\beta=2.43_{\pm 0.36}$, $\gamma=1.24_{\pm 0.74}$,
% Averages and 1-sigma errors: alpha= -0.168(79), beta=2.49(40), gamma=1.51(74)
$\alpha=-0.168(79)$, $\beta=2.49(40)$, $\gamma=1.51(74)$,
with ${\mathpzc n}_0^{\zeta_5}$ ensuring unit normalisation.

\smallskip

%\noindent\emph{5.$\;$Pion valence-quark DF at large-$x$} ---
\section{Pion valence-quark DF at large-$\mathbf X$}
The results in Fig.\,\ref{Fensemble} bear directly upon a longstanding controversy.  Namely \cite{Cui:2021mom}, analyses of the pion valence-quark DF, which incorporate the behaviour of the pion wave function prescribed by QCD, predict
\begin{equation}
\label{pionDFpQCD}
{\mathpzc u}^\pi(x;\zeta) \stackrel{x\simeq 1}{\sim} (1-x)^{\beta \,=\,2+\gamma(\zeta)}\,,
\end{equation}
where $\gamma(\zeta) \geq 0$ grows logarithmically with $\zeta$, expressing the physics of gluon radiation from the struck quark.  As noted above, $\gamma(\zeta_{\cal H})=0$.
Nevertheless, more than forty years after the first experiment \cite{Corden:1980xf} to deliver data relating to ${\mathpzc u}^\pi(x\simeq 1)$, the empirical status remains confused because, amongst the methods used to fit extant data, \emph{e.g}., Refs.\,\cite{Aicher:2010cb, Novikov:2020snp, Han:2020vjp, Barry:2021osv}, some return a ${\mathpzc u}^\pi$ form that violates Eq.\,\eqref{pionDFpQCD}.  Such disagreement requires that one of the following conclusions be faced:
the chosen analysis scheme is incomplete;
not all data included are a valid expression of qualities intrinsic to the pion;
or QCD, as currently understood, is not the theory of strong interactions.

\begin{figure}[t]
\centerline{\includegraphics[width=0.43\textwidth]{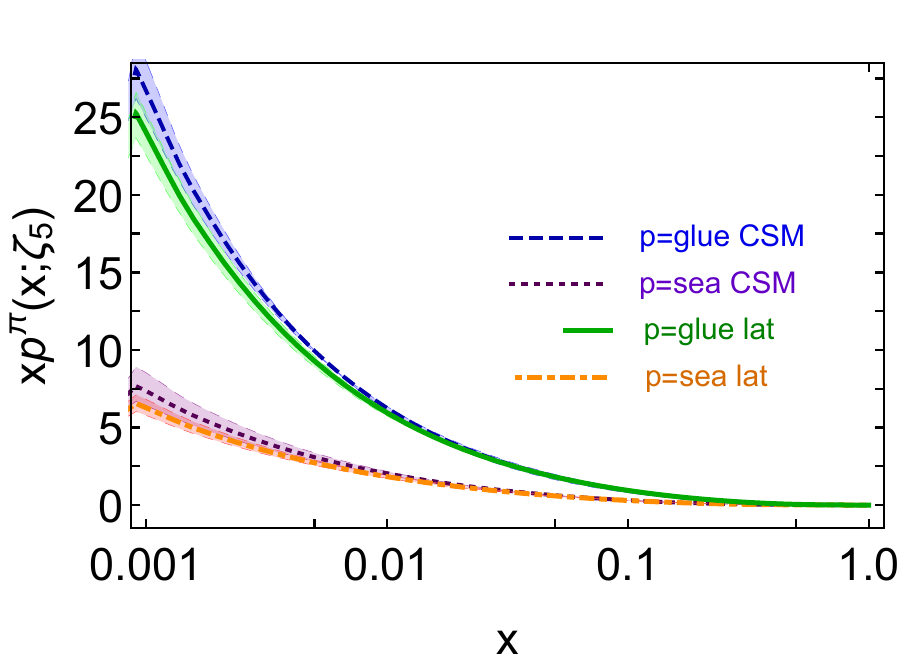}}
\caption{\label{PlotGlueSea}
Glue and sea DFs at $\zeta_5=5.2\,$GeV.  The band associated with each curve expresses consequences of the uncertainty in the valence momentum fraction:
$\langle 2 x\rangle_{{\mathpzc u}_\pi}^{\zeta_5}= 0.435(12)$; leading to
$\langle x\rangle_{{\mathpzc g}_\pi}^{\zeta_5}= 0.435(06)$,
$\langle x\rangle_{{\mathpzc S}_\pi}^{\zeta_5}= 0.125(05)$.
For comparison, CSM predictions from Refs.\,\cite{Cui:2020dlm, Cui:2020tdf, Chang:2021utv} are also drawn: in this case, $\langle 2 x\rangle_{{\mathpzc u}_\pi}^{\zeta_5}= 0.40(2)$,
$\langle x\rangle_{{\mathpzc g}_\pi}^{\zeta_5}= 0.45(1)$,
$\langle x\rangle_{{\mathpzc S}_\pi}^{\zeta_5}= 0.14(1)$.
}
\end{figure}

%\[Beta] eff = 2.28176 +/- 0.341337
Fitting the results in Fig.\,\ref{Fensemble}B on $x\in(0.9,1)$,
%Working with the results in Fig.\,\ref{Fensemble}B, fitting on $x\in(0.9,1)$,
one finds the effective value of the large-$x$ exponent: $\beta = 2.45(38)$.
Hence, the lattice simulations \cite{Joo:2019bzr, Sufian:2019bol, Alexandrou:2021mmi} yield a valence-quark DF that is consistent with Eq.\,\eqref{pionDFpQCD}.
However, the leading-order perturbative QCD analysis of data reported in Ref.\,\cite[E615]{Conway:1989fs}, which disagrees overall with the ensemble of lattice based curves, produces $\beta \approx 1.3$, contradicting Eq.\,\eqref{pionDFpQCD}.  This remains true at next-to-leading-order \cite{Novikov:2020snp, Han:2020vjp, Barry:2021osv}.
On the other hand, inclusion of soft-gluon resummation in the hard-scattering kernel produces \cite{Aicher:2010cb} the teal squares in Fig.\,\ref{Fensemble}B, which agree with the lattice-QCD ensemble and express $\beta=2.57(6)$, consistent with Eq.\,\eqref{pionDFpQCD}.
The lattice-QCD ensemble also agrees with the CSM prediction \cite{Cui:2020dlm, Cui:2020tdf, Chang:2021utv}, for which $\beta=2.81(8)$.
Recent explorations of uncertainties associated with soft-gluon resummation are briefly discussed in Appendix~\ref{SoftGlue}.
% the Supplemental Material.

Given {\sf P1}, then the results obtained above also enable prediction of the pion glue and sea DFs \cite[Sec.\,VII]{Raya:2021zrz}.  Using the central curve in Fig.\,\ref{Fensemble}A, obtained with $\rho=\rho_0 = 0.048$ in Eq.\,\eqref{EqRefSymmetric}, one arrives at the DFs in Fig.\,\ref{PlotGlueSea}.  Within uncertainties, the lattice-QCD based results calculated herein agree with the CSM predictions \cite{Cui:2020dlm, Cui:2020tdf, Chang:2021utv}.  Notably \cite{Chang:2021utv}, the CSM result for the glue DF agrees with an independent lattice determination \cite{Fan:2021bcr}; consequently, so does the result calculated herein.

%\smallskip

%\noindent\emph{6.$\;$Perspectives} ---
\section{Perspectives}
More than seventy years after discovery of the pion, Nature's most fundamental Nambu-Goldstone boson, too little is yet known about its internal structure.  This must change if the origin of nuclear-size mass-scales -- the emergence of hadron mass -- is to be understood within the Standard Model.  The proposition considered herein, \emph{viz}. that there is an effective charge which defines an evolution scheme for parton distribution functions (DFs) that is all-orders exact, has many consequences.  Amongst them, the unique definition of the hadron scale, the bounds on all Mellin moments of the valence-quark DF in pion-like systems, and the recursion relation for odd-moments, can be used to good effect, enabling, \emph{e.g}., parameter-free predictions for all pion DFs that can both benchmark existing data fitting methods and be validated using data from forthcoming experiments.  Studies are underway that test the proposition in the nucleon sector \cite{Chang:2022jri, Lu:2022cjx}.

\begin{figure}[t]
\centerline{\includegraphics[width=0.43\textwidth]{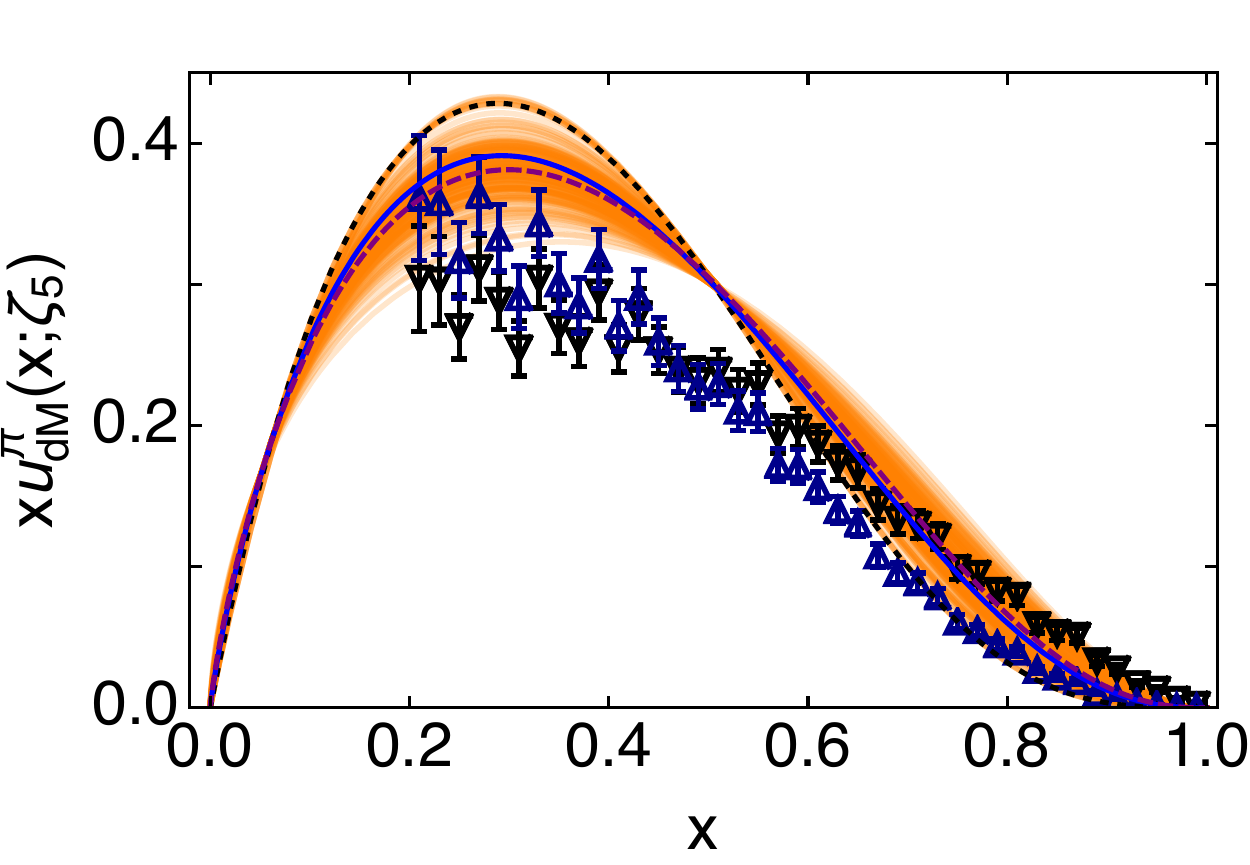}}
\caption{\label{PlotJAM}
Orange curves: ensemble of lattice-QCD based results from Fig.\,\ref{Fensemble}B.
Black down-triangles, data recorded in Ref.\,\cite[E615]{Conway:1989fs}
projected onto a  double-Mellin fit of pion structure functions; and blue up-triangles, projection onto a Mellin-Fourier fit.  (See Ref.\,\cite{Cui:2021mom}
for a  discussion.)
Both panels.  Solid blue curve: CSM prediction from Refs.\,\cite{Cui:2020dlm, Cui:2020tdf, Chang:2021utv}.
Dotted black curve: scale-free distribution.
}
\end{figure}

%
%\section*{Acknowledgments}
\smallskip
\noindent\emph{Acknowledgments}.
We are grateful for constructive comments from O.~Denisov, T.~Frederico, J.~Friedrich, C.~Mezrag, V.~Mokeev, W.-D.~Nowak, J.~Papavassiliou, C.~Quintans, G.~Salm\`e and J.~Segovia.
Work supported by:
National Natural Science Foundation of China (grant nos.\,12135007, 11805097);
Helmholtz-Zentrum Dresden-Rossendorf High Potential Programme;
Spanish Ministry of Science and Innovation (MICINN) (grant no.\ PID2019-107844GB-C22);
Junta de Andaluc{\'{\i}}a (grant nos.\ P18-FR-5057, UHU-1264517, UHU EPIT-2021);
and STRONG-2020 ``The strong interaction at the frontier of knowledge: fundamental research and applications'' which received funding from the European Union's Horizon 2020 research and innovation programme (grant no.\,824093).

\appendix

%\smallskip

%\noindent\emph{Supplemental Material} ---
\section{Soft Gluons}
\label{SoftGlue}
Uncertainties attendant upon inclusion of soft-gluon resummation in analyses of E615 data are discussed elsewhere \cite{Barry:2021osv, Cui:2021mom}.
Three different methods are compared therein.  Two may be described as Mellin-Fourier (MF) schemes \cite{Sterman:2000pt, Mukherjee:2006uu}
and yield mutually consistent results.  The Ref.\,\cite{Aicher:2010cb}
analysis is in this class.
The third is a double-Mellin (dM) approach \cite{Westmark:2017uig}.

The ensemble of lattice-QCD based results for ${\mathpzc u}^\pi(x;\zeta_5)$ is compared in Fig.\,\ref{PlotJAM}
%\ref{PlotJAM}
with reanalyses of data using the MF and dM methods.
The overall quantitative mismatch between the lattice-QCD based results and both sets of displayed data is explained by the fact that all data fits in Ref.\,\cite{Barry:2021osv}
store 15\% less of the pion's longitudinal light-front momentum with the valence degrees-of-freedom than modern calculations predict.
Regarding the large-$x$ exponent, the MF approach to soft-gluon resummation (blue up-triangles) yields $\beta_{\rm MF} = 2.24(7)$, agreeing with the lattice result and consistent with Eq.\,\eqref{pionDFpQCD}.
However, the value inferred using the dM scheme, $\beta_{\rm dM} = 1.54(5)$, is inconsistent with both the lattice result and Eq.\,\eqref{pionDFpQCD}.

%%\bibliographystyle{elsarticle-num-names}
%%\bibliography{../../../../CollectedBiB}

\begin{thebibliography}{78}
\providecommand{\natexlab}[1]{#1}
\providecommand{\url}[1]{\texttt{#1}}
\providecommand{\urlprefix}{URL }
\expandafter\ifx\csname urlstyle\endcsname\relax
  \providecommand{\doi}[1]{doi:\discretionary{}{}{}#1}\else
  \providecommand{\doi}[1]{doi:\discretionary{}{}{}\begingroup
  \urlstyle{rm}\url{#1}\endgroup}\fi
\providecommand{\bibinfo}[2]{#2}

\bibitem[{Boyanovsky et~al.(2006)Boyanovsky, de~Vega, and
  Schwarz}]{Boyanovsky:2006bf}
\bibinfo{author}{D.~Boyanovsky}, \bibinfo{author}{H.~J. de~Vega},
  \bibinfo{author}{D.~J. Schwarz}, \bibinfo{title}{{Phase transitions in the
  early and the present universe}}, \bibinfo{journal}{Ann. Rev. Nucl. Part.
  Sci.} \bibinfo{volume}{56} (\bibinfo{year}{2006}) \bibinfo{pages}{441--500}.

\bibitem[{Zyla et~al.(2020)}]{Zyla:2020zbs}
\bibinfo{author}{P.~Zyla}, et~al., \bibinfo{title}{{Review of Particle
  Physics}}, \bibinfo{journal}{PTEP} \bibinfo{volume}{2020}
  (\bibinfo{year}{2020}) \bibinfo{pages}{083C01}.

\bibitem[{Lane(1974)}]{Lane:1974he}
\bibinfo{author}{K.~D. Lane}, \bibinfo{title}{{Asymptotic Freedom and Goldstone
  Realization of Chiral Symmetry}}, \bibinfo{journal}{Phys. Rev. D}
  \bibinfo{volume}{10} (\bibinfo{year}{1974}) \bibinfo{pages}{2605}.

\bibitem[{Politzer(1976)}]{Politzer:1976tv}
\bibinfo{author}{H.~D. Politzer}, \bibinfo{title}{{Effective Quark Masses in
  the Chiral Limit}}, \bibinfo{journal}{Nucl. Phys. B} \bibinfo{volume}{117}
  (\bibinfo{year}{1976}) \bibinfo{pages}{397}.

\bibitem[{Marciano and Pagels(1978)}]{Marciano:1977su}
\bibinfo{author}{W.~J. Marciano}, \bibinfo{author}{H.~Pagels},
  \bibinfo{title}{{Quantum Chromodynamics: A Review}}, \bibinfo{journal}{Phys.
  Rept.} \bibinfo{volume}{36} (\bibinfo{year}{1978}) \bibinfo{pages}{137}.

\bibitem[{Aguilar et~al.(2008)Aguilar, Binosi, and
  Papavassiliou}]{Aguilar:2008xm}
\bibinfo{author}{A.~C. Aguilar}, \bibinfo{author}{D.~Binosi},
  \bibinfo{author}{J.~Papavassiliou}, \bibinfo{title}{{Gluon and ghost
  propagators in the Landau gauge: Deriving lattice results from
  Schwinger-Dyson equations}}, \bibinfo{journal}{Phys. Rev. D}
  \bibinfo{volume}{78} (\bibinfo{year}{2008}) \bibinfo{pages}{025010}.

\bibitem[{Gao et~al.(2018)Gao, Qin, Roberts, and
  Rodr{\'{\i}}guez-Quintero}]{Gao:2017uox}
\bibinfo{author}{F.~Gao}, \bibinfo{author}{S.-X. Qin}, \bibinfo{author}{C.~D.
  Roberts}, \bibinfo{author}{J.~Rodr{\'{\i}}guez-Quintero},
  \bibinfo{title}{{Locating the Gribov horizon}}, \bibinfo{journal}{Phys. Rev.
  D} \bibinfo{volume}{97} (\bibinfo{year}{2018}) \bibinfo{pages}{034010}.

\bibitem[{Roberts et~al.(2021)Roberts, Richards, Horn, and
  Chang}]{Roberts:2021nhw}
\bibinfo{author}{C.~D. Roberts}, \bibinfo{author}{D.~G. Richards},
  \bibinfo{author}{T.~Horn}, \bibinfo{author}{L.~Chang},
  \bibinfo{title}{{Insights into the emergence of mass from studies of pion and
  kaon structure}}, \bibinfo{journal}{Prog. Part. Nucl. Phys.}
  \bibinfo{volume}{120} (\bibinfo{year}{2021}) \bibinfo{pages}{103883}.

\bibitem[{Binosi et~al.(2017)Binosi, Mezrag, Papavassiliou, Roberts, and
  Rodr{\'i}guez-Quintero}]{Binosi:2016nme}
\bibinfo{author}{D.~Binosi}, \bibinfo{author}{C.~Mezrag},
  \bibinfo{author}{J.~Papavassiliou}, \bibinfo{author}{C.~D. Roberts},
  \bibinfo{author}{J.~Rodr{\'i}guez-Quintero},
  \bibinfo{title}{{Process-independent strong running coupling}},
  \bibinfo{journal}{Phys. Rev. D} \bibinfo{volume}{96} (\bibinfo{year}{2017})
  \bibinfo{pages}{054026}.

\bibitem[{Cui et~al.(2020{\natexlab{a}})Cui, Zhang, Binosi, de~Soto, Mezrag,
  Papavassiliou, Roberts, Rodr{\'{\i}}guez-Quintero, Segovia, and
  Zafeiropoulos}]{Cui:2019dwv}
\bibinfo{author}{Z.-F. Cui}, \bibinfo{author}{J.-L. Zhang},
  \bibinfo{author}{D.~Binosi}, \bibinfo{author}{F.~de~Soto},
  \bibinfo{author}{C.~Mezrag}, \bibinfo{author}{J.~Papavassiliou},
  \bibinfo{author}{C.~D. Roberts},
  \bibinfo{author}{J.~Rodr{\'{\i}}guez-Quintero}, \bibinfo{author}{J.~Segovia},
  \bibinfo{author}{S.~Zafeiropoulos}, \bibinfo{title}{{Effective charge from
  lattice QCD}}, \bibinfo{journal}{Chin. Phys. C} \bibinfo{volume}{44}
  (\bibinfo{year}{2020}{\natexlab{a}}) \bibinfo{pages}{083102}.

\bibitem[{Roberts(2017)}]{Roberts:2016vyn}
\bibinfo{author}{C.~D. Roberts}, \bibinfo{title}{{Perspective on the origin of
  hadron masses}}, \bibinfo{journal}{Few Body Syst.} \bibinfo{volume}{58}
  (\bibinfo{year}{2017}) \bibinfo{pages}{5}.

\bibitem[{Roberts et~al.(1992)Roberts, Williams, and Krein}]{Krein:1990sf}
\bibinfo{author}{C.~D. Roberts}, \bibinfo{author}{A.~G. Williams},
  \bibinfo{author}{G.~Krein}, \bibinfo{title}{{On the implications of
  confinement}}, \bibinfo{journal}{Int. J. Mod. Phys. A} \bibinfo{volume}{7}
  (\bibinfo{year}{1992}) \bibinfo{pages}{5607--5624}.

\bibitem[{Binosi and Tripolt(2020)}]{Binosi:2019ecz}
\bibinfo{author}{D.~Binosi}, \bibinfo{author}{R.-A. Tripolt},
  \bibinfo{title}{{Spectral functions of confined particles}},
  \bibinfo{journal}{Phys. Lett. B} \bibinfo{volume}{801} (\bibinfo{year}{2020})
  \bibinfo{pages}{135171}.

\bibitem[{Eichmann et~al.(2016)Eichmann, Sanchis-Alepuz, Williams, Alkofer, and
  Fischer}]{Eichmann:2016yit}
\bibinfo{author}{G.~Eichmann}, \bibinfo{author}{H.~Sanchis-Alepuz},
  \bibinfo{author}{R.~Williams}, \bibinfo{author}{R.~Alkofer},
  \bibinfo{author}{C.~S. Fischer}, \bibinfo{title}{{Baryons as relativistic
  three-quark bound states}}, \bibinfo{journal}{Prog. Part. Nucl. Phys.}
  \bibinfo{volume}{91} (\bibinfo{year}{2016}) \bibinfo{pages}{1--100}.

\bibitem[{Fischer(2019)}]{Fischer:2018sdj}
\bibinfo{author}{C.~S. Fischer}, \bibinfo{title}{{QCD at finite temperature and
  chemical potential from Dyson--Schwinger equations}}, \bibinfo{journal}{Prog.
  Part. Nucl. Phys.} \bibinfo{volume}{105} (\bibinfo{year}{2019})
  \bibinfo{pages}{1--60}.

\bibitem[{Qin and Roberts(2020)}]{Qin:2020rad}
\bibinfo{author}{S.-X. Qin}, \bibinfo{author}{C.~D. Roberts},
  \bibinfo{title}{{Impressions of the Continuum Bound State Problem in QCD}},
  \bibinfo{journal}{Chin. Phys. Lett.}
  \bibinfo{volume}{37}~(\bibinfo{number}{12}) (\bibinfo{year}{2020})
  \bibinfo{pages}{121201}.

\bibitem[{Roberts(2020)}]{Roberts:2020hiw}
\bibinfo{author}{C.~D. Roberts}, \bibinfo{title}{{Empirical Consequences of
  Emergent Mass}}, \bibinfo{journal}{Symmetry} \bibinfo{volume}{12}
  (\bibinfo{year}{2020}) \bibinfo{pages}{1468}.

\bibitem[{Horn and Roberts(2016)}]{Horn:2016rip}
\bibinfo{author}{T.~Horn}, \bibinfo{author}{C.~D. Roberts},
  \bibinfo{title}{{The pion: an enigma within the Standard Model}},
  \bibinfo{journal}{J. Phys. G.} \bibinfo{volume}{43} (\bibinfo{year}{2016})
  \bibinfo{pages}{073001}.

\bibitem[{Maris and Roberts(1997)}]{Maris:1997tm}
\bibinfo{author}{P.~Maris}, \bibinfo{author}{C.~D. Roberts},
  \bibinfo{title}{{{$\pi$} and {$K$} meson Bethe-Salpeter amplitudes}},
  \bibinfo{journal}{Phys. Rev. C} \bibinfo{volume}{56} (\bibinfo{year}{1997})
  \bibinfo{pages}{3369--3383}.

\bibitem[{H{\"o}ll et~al.(2004)H{\"o}ll, Krassnigg, and Roberts}]{Holl:2004fr}
\bibinfo{author}{A.~H{\"o}ll}, \bibinfo{author}{A.~Krassnigg},
  \bibinfo{author}{C.~D. Roberts}, \bibinfo{title}{{Pseudoscalar meson radial
  excitations}}, \bibinfo{journal}{Phys. Rev. C} \bibinfo{volume}{70}
  (\bibinfo{year}{2004}) \bibinfo{pages}{042203(R)}.

\bibitem[{Qin et~al.(2014)Qin, Roberts, and Schmidt}]{Qin:2014vya}
\bibinfo{author}{S.-X. Qin}, \bibinfo{author}{C.~D. Roberts},
  \bibinfo{author}{S.~M. Schmidt}, \bibinfo{title}{{Ward-Green-Takahashi
  identities and the axial-vector vertex}}, \bibinfo{journal}{Phys. Lett. B}
  \bibinfo{volume}{733} (\bibinfo{year}{2014}) \bibinfo{pages}{202--208}.

\bibitem[{Williams et~al.(2016)Williams, Fischer, and
  Heupel}]{Williams:2015cvx}
\bibinfo{author}{R.~Williams}, \bibinfo{author}{C.~S. Fischer},
  \bibinfo{author}{W.~Heupel}, \bibinfo{title}{{Light mesons in QCD and
  unquenching effects from the 3PI effective action}}, \bibinfo{journal}{Phys.
  Rev. D} \bibinfo{volume}{93} (\bibinfo{year}{2016}) \bibinfo{pages}{034026}.

\bibitem[{Binosi et~al.(2016)Binosi, Chang, Qin, Papavassiliou, and
  Roberts}]{Binosi:2016rxz}
\bibinfo{author}{D.~Binosi}, \bibinfo{author}{L.~Chang}, \bibinfo{author}{S.-X.
  Qin}, \bibinfo{author}{J.~Papavassiliou}, \bibinfo{author}{C.~D. Roberts},
  \bibinfo{title}{{Symmetry preserving truncations of the gap and
  Bethe-Salpeter equations}}, \bibinfo{journal}{Phys. Rev. D}
  \bibinfo{volume}{93} (\bibinfo{year}{2016}) \bibinfo{pages}{096010}.

\bibitem[{Qin and Roberts(2021)}]{Qin:2020jig}
\bibinfo{author}{S.-X. Qin}, \bibinfo{author}{C.~D. Roberts},
  \bibinfo{title}{{Resolving the Bethe-Salpeter kernel}},
  \bibinfo{journal}{Chin. Phys. Lett. \emph{Express}}
  \bibinfo{volume}{38}~(\bibinfo{number}{7}) (\bibinfo{year}{2021})
  \bibinfo{pages}{071201}.

\bibitem[{Carman et~al.(2020)Carman, Joo, and Mokeev}]{Carman:2020qmb}
\bibinfo{author}{D.~Carman}, \bibinfo{author}{K.~Joo},
  \bibinfo{author}{V.~Mokeev}, \bibinfo{title}{{Strong QCD Insights from
  Excited Nucleon Structure Studies with CLAS and CLAS12}},
  \bibinfo{journal}{Few Body Syst.} \bibinfo{volume}{61} (\bibinfo{year}{2020})
  \bibinfo{pages}{29}.

\bibitem[{Brodsky et~al.(2020)}]{Brodsky:2020vco}
\bibinfo{author}{S.~J. Brodsky}, et~al., \bibinfo{title}{{Strong QCD from
  Hadron Structure Experiments}}, \bibinfo{journal}{Intern. J. Mod. Phys. E}
  \bibinfo{volume}{124} (\bibinfo{year}{2020}) \bibinfo{pages}{2030006}.

\bibitem[{Barabanov et~al.(2021)}]{Barabanov:2020jvn}
\bibinfo{author}{M.~Y. Barabanov}, et~al., \bibinfo{title}{{Diquark
  Correlations in Hadron Physics: Origin, Impact and Evidence}},
  \bibinfo{journal}{Prog. Part. Nucl. Phys.} \bibinfo{volume}{116}
  (\bibinfo{year}{2021}) \bibinfo{pages}{103835}.

\bibitem[{{C. Keppel, B. Wojtsekhowski, P. King, D. Dutta, J. Annand, J. Zhang
  \emph{et al}.}(2015)}]{JlabTDIS1}
\bibinfo{author}{{C. Keppel, B. Wojtsekhowski, P. King, D. Dutta, J. Annand, J.
  Zhang \emph{et al}.}}, \bibinfo{title}{Measurement of Tagged Deep Inelastic
  Scattering (\mbox{TDIS})} \bibinfo{note}{\mbox{ }approved Jefferson Lab
  experiment E12-15-006}.

\bibitem[{{K. Park, R. Montgomery, T. Horn \emph{et al}.}(2015)}]{JlabTDIS2}
\bibinfo{author}{{K. Park, R. Montgomery, T. Horn \emph{et al}.}},
  \bibinfo{title}{Measurement of Kaon Structure Function through Tagged Deep
  Inelastic Scattering (\mbox{TDIS})} \bibinfo{note}{\mbox{ }approved Jefferson
  Lab experiment C12-15-006A.}

\bibitem[{Adams et~al.(2018)}]{Adams:2018pwt}
\bibinfo{author}{B.~Adams}, et~al., \bibinfo{title}{{Letter of Intent: A New
  QCD facility at the M2 beam line of the CERN SPS (COMPASS++/AMBER) --
  arXiv:1808.00848 [hep-ex]$\!$}} .

\bibitem[{Aguilar et~al.(2019)}]{Aguilar:2019teb}
\bibinfo{author}{A.~C. Aguilar}, et~al., \bibinfo{title}{{Pion and Kaon
  Structure at the Electron-Ion Collider}}, \bibinfo{journal}{Eur. Phys. J. A}
  \bibinfo{volume}{55} (\bibinfo{year}{2019}) \bibinfo{pages}{190}.

\bibitem[{Chen et~al.(2020)Chen, Guo, Roberts, and Wang}]{Chen:2020ijn}
\bibinfo{author}{X.~Chen}, \bibinfo{author}{F.-K. Guo}, \bibinfo{author}{C.~D.
  Roberts}, \bibinfo{author}{R.~Wang}, \bibinfo{title}{{Selected Science
  Opportunities for the EicC}}, \bibinfo{journal}{Few Body Syst.}
  \bibinfo{volume}{61} (\bibinfo{year}{2020}) \bibinfo{pages}{43}.

\bibitem[{Anderle et~al.(2021)}]{Anderle:2021wcy}
\bibinfo{author}{D.~P. Anderle}, et~al., \bibinfo{title}{{Electron-ion collider
  in China}}, \bibinfo{journal}{Front. Phys. (Beijing)}
  \bibinfo{volume}{16}~(\bibinfo{number}{6}) (\bibinfo{year}{2021})
  \bibinfo{pages}{64701}.

\bibitem[{Arrington et~al.(2021)}]{Arrington:2021biu}
\bibinfo{author}{J.~Arrington}, et~al., \bibinfo{title}{{Revealing the
  structure of light pseudoscalar mesons at the electron\textendash{}ion
  collider}}, \bibinfo{journal}{J. Phys. G} \bibinfo{volume}{48}
  (\bibinfo{year}{2021}) \bibinfo{pages}{075106}.

\bibitem[{Holt and Roberts(2010)}]{Holt:2010vj}
\bibinfo{author}{R.~J. Holt}, \bibinfo{author}{C.~D. Roberts},
  \bibinfo{title}{{Distribution Functions of the Nucleon and Pion in the
  Valence Region}}, \bibinfo{journal}{Rev. Mod. Phys.} \bibinfo{volume}{82}
  (\bibinfo{year}{2010}) \bibinfo{pages}{2991--3044}.

\bibitem[{Roberts and Williams(1994)}]{Roberts:1994dr}
\bibinfo{author}{C.~D. Roberts}, \bibinfo{author}{A.~G. Williams},
  \bibinfo{title}{{Dyson-Schwinger equations and their application to hadronic
  physics}}, \bibinfo{journal}{Prog. Part. Nucl. Phys.} \bibinfo{volume}{33}
  (\bibinfo{year}{1994}) \bibinfo{pages}{477--575}.

\bibitem[{Corden et~al.(1980)}]{Corden:1980xf}
\bibinfo{author}{M.~Corden}, et~al., \bibinfo{title}{{Production of Muon Pairs
  in the Continuum Region by 39.5-{GeV}/$c \pi^\pm$, $K^\pm$, $p$ and $\bar{p}$
  Beams Incident on a Tungsten Target}}, \bibinfo{journal}{Phys. Lett. B}
  \bibinfo{volume}{96} (\bibinfo{year}{1980}) \bibinfo{pages}{417--421}.

\bibitem[{Badier et~al.(1983)}]{Badier:1983mj}
\bibinfo{author}{J.~Badier}, et~al., \bibinfo{title}{{Experimental
  determination of the {$\pi$}-meson structure functions by the Drell-Yan
  mechanism}}, \bibinfo{journal}{Z. Phys. C} \bibinfo{volume}{18}
  (\bibinfo{year}{1983}) \bibinfo{pages}{281}.

\bibitem[{Betev et~al.(1985)}]{Betev:1985pg}
\bibinfo{author}{B.~Betev}, et~al., \bibinfo{title}{{Observation of anomalous
  scaling violation in muon pair production by 194-GeV/c {$\pi$}-tungsten
  interactions}}, \bibinfo{journal}{Z. Phys. C} \bibinfo{volume}{28}
  (\bibinfo{year}{1985}) \bibinfo{pages}{15}.

\bibitem[{Conway et~al.(1989)}]{Conway:1989fs}
\bibinfo{author}{J.~S. Conway}, et~al., \bibinfo{title}{{Experimental study of
  muon pairs produced by 252-GeV pions on tungsten}}, \bibinfo{journal}{Phys.
  Rev. D} \bibinfo{volume}{39} (\bibinfo{year}{1989}) \bibinfo{pages}{92--122}.

\bibitem[{Chang and Roberts(2021)}]{Chang:2021utv}
\bibinfo{author}{L.~Chang}, \bibinfo{author}{C.~D. Roberts},
  \bibinfo{title}{{Regarding the distribution of glue in the pion}},
  \bibinfo{journal}{Chin. Phys. Lett.}
  \bibinfo{volume}{38}~(\bibinfo{number}{8}) (\bibinfo{year}{2021})
  \bibinfo{pages}{081101}.

\bibitem[{Cui et~al.(2022)Cui, Ding, Morgado, Raya, Binosi, Chang,
  Papavassiliou, Roberts, Rodr\'\i{}guez-Quintero, and Schmidt}]{Cui:2021mom}
\bibinfo{author}{Z.~F. Cui}, \bibinfo{author}{M.~Ding}, \bibinfo{author}{J.~M.
  Morgado}, \bibinfo{author}{K.~Raya}, \bibinfo{author}{D.~Binosi},
  \bibinfo{author}{L.~Chang}, \bibinfo{author}{J.~Papavassiliou},
  \bibinfo{author}{C.~D. Roberts},
  \bibinfo{author}{J.~Rodr\'\i{}guez-Quintero}, \bibinfo{author}{S.~M.
  Schmidt}, \bibinfo{title}{{Concerning pion parton distributions}},
  \bibinfo{journal}{Eur. Phys. J. A} \bibinfo{volume}{58}~(\bibinfo{number}{1})
  (\bibinfo{year}{2022}) \bibinfo{pages}{10}.

\bibitem[{'t~Hooft(1974)}]{tHooft:1974pnl}
\bibinfo{author}{G.~'t~Hooft}, \bibinfo{title}{{A Two-Dimensional Model for
  Mesons}}, \bibinfo{journal}{Nucl. Phys. B} \bibinfo{volume}{75}
  (\bibinfo{year}{1974}) \bibinfo{pages}{461--470}.

\bibitem[{Chang et~al.(2013)Chang, Cloet, Cobos-Martinez, Roberts, Schmidt, and
  Tandy}]{Chang:2013pq}
\bibinfo{author}{L.~Chang}, \bibinfo{author}{I.~C. Cloet},
  \bibinfo{author}{J.~J. Cobos-Martinez}, \bibinfo{author}{C.~D. Roberts},
  \bibinfo{author}{S.~M. Schmidt}, \bibinfo{author}{P.~C. Tandy},
  \bibinfo{title}{{Imaging dynamical chiral symmetry breaking: pion wave
  function on the light front}}, \bibinfo{journal}{Phys. Rev. Lett.}
  \bibinfo{volume}{110} (\bibinfo{year}{2013}) \bibinfo{pages}{132001}.

\bibitem[{Diehl et~al.(2001)Diehl, Feldmann, Jakob, and Kroll}]{Diehl:2000xz}
\bibinfo{author}{M.~Diehl}, \bibinfo{author}{T.~Feldmann},
  \bibinfo{author}{R.~Jakob}, \bibinfo{author}{P.~Kroll}, \bibinfo{title}{{The
  Overlap representation of skewed quark and gluon distributions}},
  \bibinfo{journal}{Nucl. Phys. B} \bibinfo{volume}{596} (\bibinfo{year}{2001})
  \bibinfo{pages}{33--65}.

\bibitem[{Brodsky and Lepage(1979)}]{Brodsky:1979gy}
\bibinfo{author}{S.~J. Brodsky}, \bibinfo{author}{G.~P. Lepage},
  \bibinfo{title}{{Perturbative Quantum Chromodynamics}},
  \bibinfo{journal}{Prog. Math. Phys.} \bibinfo{volume}{4}
  (\bibinfo{year}{1979}) \bibinfo{pages}{255--422}.

\bibitem[{Dokshitzer(1977)}]{Dokshitzer:1977sg}
\bibinfo{author}{Y.~L. Dokshitzer}, \bibinfo{title}{Calculation of the
  Structure Functions for Deep Inelastic Scattering and e+ e- Annihilation by
  Perturbation Theory in Quantum Chromodynamics. ({\mbox {I}n {R}ussian})},
  \bibinfo{journal}{Sov. Phys. JETP} \bibinfo{volume}{46}
  (\bibinfo{year}{1977}) \bibinfo{pages}{641--653}.

\bibitem[{Gribov and Lipatov(1971)}]{Gribov:1971zn}
\bibinfo{author}{V.~N. Gribov}, \bibinfo{author}{L.~N. Lipatov},
  \bibinfo{title}{{Deep inelastic electron scattering in perturbation theory}},
  \bibinfo{journal}{Phys. Lett. B} \bibinfo{volume}{37} (\bibinfo{year}{1971})
  \bibinfo{pages}{78--80}.

\bibitem[{Lipatov(1975)}]{Lipatov:1974qm}
\bibinfo{author}{L.~N. Lipatov}, \bibinfo{title}{{The parton model and
  perturbation theory}}, \bibinfo{journal}{Sov. J. Nucl. Phys.}
  \bibinfo{volume}{20} (\bibinfo{year}{1975}) \bibinfo{pages}{94--102}.

\bibitem[{Altarelli and Parisi(1977)}]{Altarelli:1977zs}
\bibinfo{author}{G.~Altarelli}, \bibinfo{author}{G.~Parisi},
  \bibinfo{title}{{Asymptotic Freedom in Parton Language}},
  \bibinfo{journal}{Nucl. Phys. B} \bibinfo{volume}{126} (\bibinfo{year}{1977})
  \bibinfo{pages}{298--318}.

\bibitem[{Ding et~al.(2020{\natexlab{a}})Ding, Raya, Binosi, Chang, Roberts,
  and Schmidt}]{Ding:2019qlr}
\bibinfo{author}{M.~Ding}, \bibinfo{author}{K.~Raya},
  \bibinfo{author}{D.~Binosi}, \bibinfo{author}{L.~Chang},
  \bibinfo{author}{C.~D. Roberts}, \bibinfo{author}{S.~M. Schmidt},
  \bibinfo{title}{{Drawing insights from pion parton distributions}},
  \bibinfo{journal}{Chin. Phys. C (Lett.)} \bibinfo{volume}{44}
  (\bibinfo{year}{2020}{\natexlab{a}}) \bibinfo{pages}{031002}.

\bibitem[{Ding et~al.(2020{\natexlab{b}})Ding, Raya, Binosi, Chang, Roberts,
  and Schmidt}]{Ding:2019lwe}
\bibinfo{author}{M.~Ding}, \bibinfo{author}{K.~Raya},
  \bibinfo{author}{D.~Binosi}, \bibinfo{author}{L.~Chang},
  \bibinfo{author}{C.~D. Roberts}, \bibinfo{author}{S.~M. Schmidt},
  \bibinfo{title}{{Symmetry, symmetry breaking, and pion parton
  distributions}}, \bibinfo{journal}{Phys. Rev. D}
  \bibinfo{volume}{101}~(\bibinfo{number}{5})
  (\bibinfo{year}{2020}{\natexlab{b}}) \bibinfo{pages}{054014}.

\bibitem[{Cui et~al.(2021)Cui, Ding, Gao, Raya, Binosi, Chang, Roberts,
  Rodr\'{\i}guez-Quintero, and Schmidt}]{Cui:2020dlm}
\bibinfo{author}{Z.-F. Cui}, \bibinfo{author}{M.~Ding},
  \bibinfo{author}{F.~Gao}, \bibinfo{author}{K.~Raya},
  \bibinfo{author}{D.~Binosi}, \bibinfo{author}{L.~Chang},
  \bibinfo{author}{C.~D. Roberts},
  \bibinfo{author}{J.~Rodr\'{\i}guez-Quintero}, \bibinfo{author}{S.~M.
  Schmidt}, \bibinfo{title}{{Higgs modulation of emergent mass as revealed in
  kaon and pion parton distributions}}, \bibinfo{journal}{Eur. Phys. J. A
  (Lett.)} \bibinfo{volume}{57}~(\bibinfo{number}{1}) (\bibinfo{year}{2021})
  \bibinfo{pages}{5}.

\bibitem[{Cui et~al.(2020{\natexlab{b}})Cui, Ding, Gao, Raya, Binosi, Chang,
  Roberts, Rodr\'{\i}guez-Quintero, and Schmidt}]{Cui:2020tdf}
\bibinfo{author}{Z.-F. Cui}, \bibinfo{author}{M.~Ding},
  \bibinfo{author}{F.~Gao}, \bibinfo{author}{K.~Raya},
  \bibinfo{author}{D.~Binosi}, \bibinfo{author}{L.~Chang},
  \bibinfo{author}{C.~D. Roberts},
  \bibinfo{author}{J.~Rodr\'{\i}guez-Quintero}, \bibinfo{author}{S.~M.
  Schmidt}, \bibinfo{title}{{Kaon and pion parton distributions}},
  \bibinfo{journal}{Eur. Phys. J. C} \bibinfo{volume}{80}
  (\bibinfo{year}{2020}{\natexlab{b}}) \bibinfo{pages}{1064}.

\bibitem[{Zhang et~al.(2021)Zhang, Cui, Ping, and Roberts}]{Zhang:2020ecj}
\bibinfo{author}{J.-L. Zhang}, \bibinfo{author}{Z.-F. Cui},
  \bibinfo{author}{J.~Ping}, \bibinfo{author}{C.~D. Roberts},
  \bibinfo{title}{{Contact interaction analysis of pion GTMDs}},
  \bibinfo{journal}{Eur. Phys. J. C} \bibinfo{volume}{81}~(\bibinfo{number}{1})
  (\bibinfo{year}{2021}) \bibinfo{pages}{6}.

\bibitem[{Grunberg(1984)}]{Grunberg:1982fw}
\bibinfo{author}{G.~Grunberg}, \bibinfo{title}{{Renormalization Scheme
  Independent QCD and QED: The Method of Effective Charges}},
  \bibinfo{journal}{Phys. Rev. D} \bibinfo{volume}{29} (\bibinfo{year}{1984})
  \bibinfo{pages}{2315}.

\bibitem[{Grunberg(1989)}]{Grunberg:1989xf}
\bibinfo{author}{G.~Grunberg}, \bibinfo{title}{{On Some Ambiguities in the
  Method of Effective Charges}}, \bibinfo{journal}{Phys. Rev. D}
  \bibinfo{volume}{40} (\bibinfo{year}{1989}) \bibinfo{pages}{680}.

\bibitem[{Dokshitzer(1998)}]{Dokshitzer:1998nz}
\bibinfo{author}{Y.~L. Dokshitzer}, \bibinfo{title}{{\emph{Perturbative QCD
  theory (includes our knowledge of \mbox{$\alpha(s)$})} - hep-ph/9812252}},
  in: \bibinfo{booktitle}{{High-energy physics. Proceedings, 29th International
  Conference, ICHEP'98, Vancouver, Canada, July 23-29, 1998. Vol. 1, 2}},
  \bibinfo{pages}{305--324}, \bibinfo{year}{1998}.

\bibitem[{Deur et~al.(2016)Deur, Brodsky, and de~Teramond}]{Deur:2016tte}
\bibinfo{author}{A.~Deur}, \bibinfo{author}{S.~J. Brodsky},
  \bibinfo{author}{G.~F. de~Teramond}, \bibinfo{title}{{The QCD Running
  Coupling}}, \bibinfo{journal}{Prog. Part. Nucl. Phys.} \bibinfo{volume}{90}
  (\bibinfo{year}{2016}) \bibinfo{pages}{1--74}.

\bibitem[{Brodsky et~al.(1995)Brodsky, Burkardt, and Schmidt}]{Brodsky:1994kg}
\bibinfo{author}{S.~J. Brodsky}, \bibinfo{author}{M.~Burkardt},
  \bibinfo{author}{I.~Schmidt}, \bibinfo{title}{{Perturbative QCD constraints
  on the shape of polarized quark and gluon distributions}},
  \bibinfo{journal}{Nucl. Phys. B} \bibinfo{volume}{441} (\bibinfo{year}{1995})
  \bibinfo{pages}{197--214}.

\bibitem[{Yuan(2004)}]{Yuan:2003fs}
\bibinfo{author}{F.~Yuan}, \bibinfo{title}{{Generalized parton distributions at
  $x \to 1$}}, \bibinfo{journal}{Phys. Rev. D} \bibinfo{volume}{69}
  (\bibinfo{year}{2004}) \bibinfo{pages}{051501}.

\bibitem[{Ball et~al.(2017)}]{NNPDF:2017mvq}
\bibinfo{author}{R.~D. Ball}, et~al., \bibinfo{title}{{Parton distributions
  from high-precision collider data}}, \bibinfo{journal}{Eur. Phys. J. C}
  \bibinfo{volume}{77}~(\bibinfo{number}{10}) (\bibinfo{year}{2017})
  \bibinfo{pages}{663}.

\bibitem[{Hou et~al.(2021)}]{Hou:2019efy}
\bibinfo{author}{T.-J. Hou}, et~al., \bibinfo{title}{{New CTEQ global analysis
  of quantum chromodynamics with high-precision data from the LHC}},
  \bibinfo{journal}{Phys. Rev. D} \bibinfo{volume}{103}~(\bibinfo{number}{1})
  (\bibinfo{year}{2021}) \bibinfo{pages}{014013}.

\bibitem[{Diehl and Stienemeier(2020)}]{Diehl:2019fsz}
\bibinfo{author}{M.~Diehl}, \bibinfo{author}{P.~Stienemeier},
  \bibinfo{title}{{Gluons and sea quarks in the proton at low scales}},
  \bibinfo{journal}{Eur. Phys. J. Plus}
  \bibinfo{volume}{135}~(\bibinfo{number}{2}) (\bibinfo{year}{2020})
  \bibinfo{pages}{211}.

\bibitem[{Jo\'o et~al.(2019)Jo\'o, Karpie, Orginos, Radyushkin, Richards,
  Sufian, and Zafeiropoulos}]{Joo:2019bzr}
\bibinfo{author}{B.~Jo\'o}, \bibinfo{author}{J.~Karpie},
  \bibinfo{author}{K.~Orginos}, \bibinfo{author}{A.~V. Radyushkin},
  \bibinfo{author}{D.~G. Richards}, \bibinfo{author}{R.~S. Sufian},
  \bibinfo{author}{S.~Zafeiropoulos}, \bibinfo{title}{{Pion valence structure
  from Ioffe-time parton pseudodistribution functions}},
  \bibinfo{journal}{Phys. Rev. D} \bibinfo{volume}{100} (\bibinfo{year}{2019})
  \bibinfo{pages}{114512}.

\bibitem[{Sufian et~al.(2019)Sufian, Karpie, Egerer, Orginos, Qiu, and
  Richards}]{Sufian:2019bol}
\bibinfo{author}{R.~S. Sufian}, \bibinfo{author}{J.~Karpie},
  \bibinfo{author}{C.~Egerer}, \bibinfo{author}{K.~Orginos},
  \bibinfo{author}{J.-W. Qiu}, \bibinfo{author}{D.~G. Richards},
  \bibinfo{title}{{Pion Valence Quark Distribution from Matrix Element
  Calculated in Lattice QCD}}, \bibinfo{journal}{Phys. Rev. D}
  \bibinfo{volume}{99} (\bibinfo{year}{2019}) \bibinfo{pages}{074507}.

\bibitem[{Alexandrou et~al.(2021)Alexandrou, Bacchio, Cloet, Constantinou,
  Hadjiyiannakou, Koutsou, and Lauer}]{Alexandrou:2021mmi}
\bibinfo{author}{C.~Alexandrou}, \bibinfo{author}{S.~Bacchio},
  \bibinfo{author}{I.~Cloet}, \bibinfo{author}{M.~Constantinou},
  \bibinfo{author}{K.~Hadjiyiannakou}, \bibinfo{author}{G.~Koutsou},
  \bibinfo{author}{C.~Lauer}, \bibinfo{title}{{Pion and kaon $\langle
  x^3\rangle$ from lattice QCD and PDF reconstruction from Mellin moments}},
  \bibinfo{journal}{Phys. Rev. D} \bibinfo{volume}{104}~(\bibinfo{number}{5})
  (\bibinfo{year}{2021}) \bibinfo{pages}{054504}.

\bibitem[{Raya et~al.(2022)Raya, Cui, Chang, Morgado, Roberts, and
  Rodr{\'{\i}}guez-Quintero}]{Raya:2021zrz}
\bibinfo{author}{K.~Raya}, \bibinfo{author}{Z.-F. Cui},
  \bibinfo{author}{L.~Chang}, \bibinfo{author}{J.-M. Morgado},
  \bibinfo{author}{C.~D. Roberts},
  \bibinfo{author}{J.~Rodr{\'{\i}}guez-Quintero}, \bibinfo{title}{{Revealing
  pion and kaon structure via generalised parton distributions}},
  \bibinfo{journal}{Chin. Phys. C} \bibinfo{volume}{46} (\bibinfo{year}{2022})
  \bibinfo{pages}{013107}.

\bibitem[{Aicher et~al.(2010)Aicher, Sch{\"a}fer, and
  Vogelsang}]{Aicher:2010cb}
\bibinfo{author}{M.~Aicher}, \bibinfo{author}{A.~Sch{\"a}fer},
  \bibinfo{author}{W.~Vogelsang}, \bibinfo{title}{{Soft-Gluon Resummation and
  the Valence Parton Distribution Function of the Pion}},
  \bibinfo{journal}{Phys.\ Rev.\ Lett.} \bibinfo{volume}{105}
  (\bibinfo{year}{2010}) \bibinfo{pages}{252003}.

\bibitem[{Novikov et~al.(2020)}]{Novikov:2020snp}
\bibinfo{author}{I.~Novikov}, et~al., \bibinfo{title}{{Parton Distribution
  Functions of the Charged Pion Within The xFitter Framework}},
  \bibinfo{journal}{Phys. Rev. D} \bibinfo{volume}{102} (\bibinfo{year}{2020})
  \bibinfo{pages}{014040}.

\bibitem[{Han et~al.(2021)Han, Xie, Wang, and Chen}]{Han:2020vjp}
\bibinfo{author}{C.~Han}, \bibinfo{author}{G.~Xie}, \bibinfo{author}{R.~Wang},
  \bibinfo{author}{X.~Chen}, \bibinfo{title}{{An Analysis of Parton
  Distribution Functions of the Pion and the Kaon with the Maximum Entropy
  Input}}, \bibinfo{journal}{Eur. Phys. J. C} \bibinfo{volume}{81}
  (\bibinfo{year}{2021}) \bibinfo{pages}{302}.

\bibitem[{Barry et~al.(2021)Barry, Ji, Sato, and Melnitchouk}]{Barry:2021osv}
\bibinfo{author}{P.~C. Barry}, \bibinfo{author}{C.-R. Ji},
  \bibinfo{author}{N.~Sato}, \bibinfo{author}{W.~Melnitchouk},
  \bibinfo{title}{{Global QCD Analysis of Pion Parton Distributions with
  Threshold Resummation}}, \bibinfo{journal}{Phys. Rev. Lett.}
  \bibinfo{volume}{127}~(\bibinfo{number}{23}) (\bibinfo{year}{2021})
  \bibinfo{pages}{232001}.

\bibitem[{Fan and Lin(2021)}]{Fan:2021bcr}
\bibinfo{author}{Z.~Fan}, \bibinfo{author}{H.-W. Lin}, \bibinfo{title}{{Gluon
  parton distribution of the pion from lattice QCD}}, \bibinfo{journal}{Phys.
  Lett. B} \bibinfo{volume}{823} (\bibinfo{year}{2021})
  \bibinfo{pages}{136778}.

\bibitem[{Chang et~al.(2022)Chang, Gao, and Roberts}]{Chang:2022jri}
\bibinfo{author}{L.~Chang}, \bibinfo{author}{F.~Gao}, \bibinfo{author}{C.~D.
  Roberts}, \bibinfo{title}{{Parton distributions of light quarks and
  antiquarks in the proton -- arXiv:2201.07870 [hep-ph]}},
  \bibinfo{journal}{Phys. Lett. B} \bibinfo{volume}{829} (\bibinfo{year}{2022})
  \bibinfo{pages}{137078}.

\bibitem[{Lu et~al.(2022)Lu, Chang, Raya, Roberts, and
  Rodr\'\i{}guez-Quintero}]{Lu:2022cjx}
\bibinfo{author}{Y.~Lu}, \bibinfo{author}{L.~Chang}, \bibinfo{author}{K.~Raya},
  \bibinfo{author}{C.~D. Roberts},
  \bibinfo{author}{J.~Rodr\'\i{}guez-Quintero}, \bibinfo{title}{{Proton and
  pion distribution functions in counterpoint -- arXiv:2203.00753 [hep-ph]}} .

\bibitem[{Sterman and Vogelsang(2001)}]{Sterman:2000pt}
\bibinfo{author}{G.~F. Sterman}, \bibinfo{author}{W.~Vogelsang},
  \bibinfo{title}{{Threshold resummation and rapidity dependence}},
  \bibinfo{journal}{JHEP} \bibinfo{volume}{02} (\bibinfo{year}{2001})
  \bibinfo{pages}{016}.

\bibitem[{Mukherjee and Vogelsang(2006)}]{Mukherjee:2006uu}
\bibinfo{author}{A.~Mukherjee}, \bibinfo{author}{W.~Vogelsang},
  \bibinfo{title}{{Threshold resummation for W-boson production at RHIC}},
  \bibinfo{journal}{Phys. Rev. D} \bibinfo{volume}{73} (\bibinfo{year}{2006})
  \bibinfo{pages}{074005}.

\bibitem[{Westmark and Owens(2017)}]{Westmark:2017uig}
\bibinfo{author}{D.~Westmark}, \bibinfo{author}{J.~F. Owens},
  \bibinfo{title}{{Enhanced threshold resummation formalism for lepton pair
  production and its effects in the determination of parton distribution
  functions}}, \bibinfo{journal}{Phys. Rev. D} \bibinfo{volume}{95}
  (\bibinfo{year}{2017}) \bibinfo{pages}{056024}.

\end{thebibliography}

\end{document}